\documentclass[oneside,reviewcopy,10pt]{elsart}

\usepackage{graphicx}
\usepackage{subfigure}

\newcommand{\nn}{\nonumber}

\usepackage{amssymb}

\begin{document}

\begin{frontmatter}
\title{A new parallel strategy for two-dimensional incompressible
flow simulations using pseudo-spectral methods}
\author[address1]{Z. Yin},
\ead{yinzh@lsec.cc.ac.cn}
\author[address1]{Li Yuan},
\ead{lyuan@lsec.cc.ac.cn}
\author[address2,address1]{Tao Tang}
\ead{ttang@math.hkbu.edu.hk}

\address[address1]{LSEC, Institute of Computational Mathematics, Chinese
Academy of Sciences, \\ P.O. Box 2719, Beijing 100080, P.R.China}

\address[address2]{Department of Mathematics, Hong Kong Baptist
University, Hong Kong, \\ P.R. China}

\begin{abstract}
A novel parallel technique for Fourier-Galerkin pseudo-spectral
methods with applications to two-dimensional Navier-Stokes
equations and inviscid Boussinesq approximation equations is
presented. It takes the advantage of the programming structure of
the phase-shift de-aliased scheme for pseudo-spectral codes, and
combines the task-distribution strategy
[Yin, Clercx and Montgomery, Comput. Fluids, 33, 509 (2004)]
and parallelized Fast Fourier Transform scheme. The performances of
the resulting MPI Fortran90 codes with the new procedure on SGI
3800 are reported. For fixed resolution of the same problem, the
peak speed of the new scheme can be twice as fast as the old
parallel methods. The parallelized codes are used to solve some
challenging numerical problems governed by the Navier-Stokes
equations and the Boussinesq equations. Two interesting physical
problems, namely, the double-valued $\omega$-$\psi $ structure in
two-dimensional decaying turbulence and the collapse of the
bubble cap in the Boussinesq simulation, are solved by
using the proposed parallel algorithms.
\end{abstract}

\end{frontmatter}

\section{Introduction}
 The pseudo-spectral method has been very popular in the
research of highly accurate numerical simulations since the
pioneer work of Orszag and Patterson ~\cite{orsz71,patt71}. For
smooth solutions, the convergence order of the spectral methods
is higher than any algebraic power of mesh size. For a comparable
error on the uniform mesh, a much finer mesh is required for
finite difference or finite element methods. This is one of the
reasons that the spectral method has been widely used in spite of
the prosperous development of adaptive grid methods.
Some comprehensive
overviews of the various applications of spectral methods in
fluid dynamics can be found in ~\cite{canu87,boyd}.
With the fast development of supercomputers, more and more
efforts have been devoted to parallelizing spectral methods. For
challenging simulation problems such as turbulence research, many
new interesting phenomena are discovered using large scale
parallel computations ($512^3 - 4096^3$) (e.g. see
~\cite{chen93,kane03}, and a review paper ~\cite{moin98}). The
parallel schemes based on spectral methods have also been
intensively investigated over the last decade---mainly for
three-dimensional (3D) problems
~\cite{pelz91,jack91,fisc94,bris95,dmit01,iovi01,basu94,ling02}.
The main computation time for spectral methods is concentrated in
the part of the Fast Fourier Transform (FFT), around which the 3D
parallel schemes are constructed. In the following, we will
denote this parallel FFT procedure as PFFT.

For high dimensional FFT, the transpose-split method can provide
very high parallel efficiency, so most commonly used 3D parallel
schemes employ this kind of parallel FFT
~\cite{pelz91,jack91,fisc94,bris95,dmit01}.
Iovieno \textit{et al.} ~\cite{iovi01} combined the
transpose-split method and de-aliased procedure in
pseudo-spectral codes together to yield high parallel efficiency.
Based on the three time-evolution equations of the 3D
Navier-Stokes (NS) equations, Basu adopted a parallel scheme
which can only be used on a 3-processor computer ~\cite{basu94}.
Ling \textit{et al.} ~\cite{ling02} try to parallel the 3D code
by combining the 3-CPU method and the PFFT scheme, and a
comparison showed that the combined scheme is always slower than
PFFT.

In contrast to the world-wide efforts in 3D parallelization,
rather limited efforts have been devoted to parallelizing the
two-dimensional simulations ~\cite{four02,zyin04}. It is quite
common to treat the 2D parallel spectral code as a simplified
version of the 3D codes, which are very efficient only in 3D cases
(PFFT). As a result, the ratio of the communication time to the
computation time in those 2D parallel codes is relatively large,
and the parallel efficiency is much lower compared with the
corresponding 3D codes (see a short discussion about this topic in
~\cite{zyin04}).

To minimize the relative long communication time, Yin \textit{et
al.} ~\cite{zyin04} propose a parallel task distribution scheme
(PTD) in the 2D pseudo-spectral NS code. Although this scheme is
very easy to implement with a good parallel efficiency, it has
the limitation that the code can only use 2, 4, and 6 processors
to do the calculation.

In this paper, we will parallel the 2D spectral codes by
combining the PTD and PFFT schemes. The new strategy overcomes the
shortcomings of the former schemes and shows a significant
improvement in parallel efficiency. In the following, we will
denote this combined strategy as PTF scheme (parallelization
through task distribution and FFT).

The paper is organized as follows.
In Section 2, the PTF scheme is applied to solve the 2D NS
equation; the benchmark of the parallel code on SGI 3800 is
presented. We also show several long-time numerical simulations
with high resolutions, which reveal some interesting physical
phenomena of 2D decaying turbulence ~\cite{zyin03,zyin04a}. Of
course, PTF plays an important role in those long-time runs. In
section 3, the new scheme is applied to solve the 2D inviscid
Boussinesq equations. A challenging numerical problem, which is
studied previously in ~\cite{pumi92,we94,ceni01}, is investigated
with three kinds of resolutions ($1024^2$, $2048^2$, and
$4096^2$). Finally, a summary of this work is given, where the
prospects of the new scheme are also discussed.

\section{Parallel 2D pseudo-spectral code for the NS
equations}

The study of the 2D turbulence distinguishes itself from the 3D
turbulence due to its unique phenomena such as inverse energy
cascade and self-organization. In the past few decades, a
particular kind of statistical mechanics~\cite{mont74} has been
widely adopted to study the 2D freely decaying turbulence (see
~\cite{zyin03} and references therein).

As a powerful tool, direct numerical simulations (DNS) provide
some useful theory check and inspire the deeper thoughts of the
statistical theory (e.g. see ~\cite{zyin03,zyin04a}). Those
simulations normally need to last as long as 100-1000 eddy
turnover times before final states of the 2D decaying turbulence
are reached. This means that the total calculations of this kind
of 2D DNS, although has a smaller array, are more or less the same
as those of some short-term 3D DNS. For example, 20 time steps of
a 2D DNS with the resolution of $1024^2$ will take roughly the
same CPU time as one time step of a 3D DNS on a grid of $256^3$
on the same computer. On the other hand, because the time
evolution loop is impossible to be parallelized, the parallel
procedure within one time step becomes essential to improve the
performance of a 2D DNS code.

On modern supercomputers, the total CPU time involved in the 2D
DNS is not as enormous as that of the 3D DNS if the grid points
are the same in each dimension. The peak performance of the
parallel code, which is indicated by the shortest wall clock time
regardless of the number of CPUs used, becomes more important in
2D DNS. In our research, we would like to get our 2D DNS results
in 3 days using 32 CPUs rather than wait for a week using 8 CPUs,
although we only achieve a speedup of 2.33 by the fourfold
processors. As will be seen later in this paper, the significant
improvement of the peak speed is a big advantage of the PTF
scheme.

\subsection{The governing equations and pseudo-spectral methods}

The 2D incompressible NS equation are usually written as
\begin{equation}
\label{eq1}
 \frac{\partial {\rm {\bf u}}}{\partial t} + {\rm {\bf
u}} \cdot \nabla {\rm {\bf u}} = - \nabla p + \nu \Delta {\rm {\bf
u}},
\end{equation}
\begin{equation}
\label{eq2} \nabla \cdot {\rm {\bf u}} = 0,
\end{equation}
where ${\rm {\bf u}} = (\mbox{u,v})$ is the velocity, $p$ is the
pressure, and $\nu$ is the kinematic viscosity. Using the
vorticity ${\rm {\bf \omega }} = (0,0,\omega ) = \nabla \times
{\rm {\bf u}}$ and stream function $\psi $, Eq. (1) and (2) can be
rewritten as:
\begin{equation}
\label{eq3} \frac{\partial \omega }{\partial t} + {\rm {\bf u}}
\cdot \nabla \omega = \nu \Delta \omega ,
\end{equation}
\begin{equation}
\label{eq4} \Delta \psi = - \omega .
\end{equation}

The stream function is related to velocity by $\mbox{u} =
{\partial \psi } \mathord{\left/ {\vphantom {{\partial \psi }
{\partial y}}} \right. \kern-\nulldelimiterspace} {\partial y}$
and $\mbox{v} = - {\partial \psi } \mathord{\left/ {\vphantom
{{\partial \psi } {\partial x}}} \right.
\kern-\nulldelimiterspace} {\partial x}$. In the conservative
form, Eq. (3) becomes
\begin{equation}
\label{eq5} \frac{\partial \omega }{\partial t} + \nabla \cdot
(\omega {\kern 1pt} {\rm {\bf u}}) = \nu \Delta \omega.
\end{equation}

We adopted ABCN scheme to carry out the time integration, which
discretizes the nonlinear term $\nabla \cdot (\omega {\kern 1pt}
{\rm {\bf u}})$ with a 2$^{nd}$ order Adams-Bashforth scheme and
the dissipation term $\nu \Delta \omega $ with the Crank-Nicolson
scheme. The particular time-stepping used is independent of the
parallelization. More accurate schemes, such as 4$^{th}$ order
Runge-Kutta, could also be used.

When Eqs. (4) and (5) are solved by Galerkin pseudo-spectral
methods, the Fourier coefficients of $\omega $ are evaluated at
each time step. The nonlinear term $\nabla \cdot (\omega {\kern
1pt} {\rm {\bf u}})$ is obtained by being transferred back and
from the physical space with FFTs; in the meantime, the
de-aliasing procedure has to be adopted to remove the aliasing
error. There are two kinds of de-aliasing techniques available:
padding-truncation and phase-shifts ~\cite{canu87}. The FFT we
used requires the dimension size to be the power of 2, which is
faster than those FFTs that has prime factors other than 2. The
costs of two de-aliasing techniques are the same as that for our
FFTs. Therefore, in this paper we use the phase-shifts, which
reserves more high wave numbers information than that for
padding-truncation.

\begin{table}[t!]
\caption{Ten FFTs need to be calculated in each time step of NS
spectral solver.} \centering
\begin{tabular}
 {|p{12pt}|p{58pt}|p{58pt}|p{58pt}|p{58pt}|} \hline \hline & A&
B& C&
D \\
\hline 1& $\hat{\mbox{u}} \to \mbox{u}$& $\hat{\mbox{v}} \to
\mbox{v}$& $\hat{\mbox{U}} \to \mbox{U}$&
$\hat{\mbox {V}} \to \mbox{V}$ \\
\hline 2& \multicolumn{2}{|p{116pt}|}{$\hspace{0.7in} \hat {\omega
} \to \omega $} &
\multicolumn{2}{|p{116pt}|}{$\hspace{0.7in} \hat {\Omega } \to \Omega $}  \\
\hline 3& $\mbox{u}\omega \to \widehat{\left( {\mbox{u}\omega }
\right)}$& $\mbox{v}\omega \to \widehat{\left( {\mbox{v}\omega }
\right)}$& $\mbox{U}\Omega \to \widehat{\left( {\mbox{U}\Omega }
\right)}$&
$\mbox{V}\Omega \to \widehat{\left( {\mbox{V}\Omega } \right)}$ \\
\hline \hline
\end{tabular}
\end{table}

Table 1 shows the ten FFTs needed to solve Eqs. (4) and (5) in
each time step. The expression with a hat (e.g. $\hat
{u},\;\widehat{(\mbox{U}\Omega )}$) is the spectral space value
of the corresponding physical value without hat ($u,\;\left(
{\mbox{U$\Omega$ }} \right)$, etc.). The capital letters denote
the phase-shift values of the corresponding lower case variables.
For example, in a simulation with $M^2$ resolution:
\[
\Omega _{\rm {\bf j}} = \sum\limits_{\rm {\bf k}} {\hat {\omega
}_k e^{i{\rm {\bf k}} \cdot \left( {{\rm {\bf x}}_{\rm {\bf j}} +
\delta } \right)}} ,
\]
where, ${\rm {\bf j}}$ indicates the grid point, ${\rm {\bf k}}$
is the wave number, ${\rm {\bf x}} = (x,y)$, and $\delta = \left(
{\pi / M,\pi / M} \right)$. Each arrow in Table 1 indicated one
2D FFT, and the four FFTs in the third row should be calculated
after the six FFTs in the first two rows (see Fig.3 of
~\cite{zyin04} for an even clear picture).

It is worth to mention that the pseudo-spectral method needs to
evaluate twelve FFTs for the non-conservative form (Eqs. (3) and
(4))\footnote{{\small The nonlinear term in Eq. (1) and Fig.3 of
Ref ~\cite{zyin04} is inconsistently written in the
non-conservative form (}${\rm {\bf u}} \cdot \nabla \omega
${\small ), which should be in conservative form (}$\nabla \cdot
(\omega {\kern 1pt} {\rm {\bf u}})${\small ) to march the
structure of the program.}}, since the two FFTs in the 2$^{nd}$
row of Table 1 are replaced by four FFTs to transfer
$\widehat{\left( {{\partial \omega } \mathord{\left/ {\vphantom
{{\partial \omega } {\partial x}}} \right.
\kern-\nulldelimiterspace} {\partial x}} \right)}$,
$\widehat{\left( {{\partial \omega } \mathord{\left/ {\vphantom
{{\partial \omega } {\partial y}}} \right.
\kern-\nulldelimiterspace} {\partial y}} \right)}$, and their
phase-shift counterparts to physical space. (Of course, the FFTs
in the 3$^{rd}$ row need to be changed correspondingly although
no more FFT is needed.) Hence, it is a natural choice to use the
conservative formulation when we use PFFT scheme to parallel the
2D NS code.

In the following subsection, we will give a brief discussion for
the PFFT scheme. We will also introduce some symbols and analyzing
tools that will be used throughout the paper.

\subsection{A brief discussion for the PFFT scheme}

The PFFT scheme calculates the ten parallelized FFTs one after
another according to certain sequence mentioned in the previous
subsection. In our case, the parallelized FFT is the
widely-adopted transpose-split parallel scheme, which computes one
dimensional (1D) FFT in each direction combined with the data
transfer between CPUs and matrix transpose ~\cite{jack91,cchu89}.

It is observed that the research on parallel FFT is a fast
developed field (e.g. see ~\cite{swar87,cham88,pelz93,merm03}, or
a relatively complete review in ~\cite{echu00}). It is difficult
to rank the available FFTs, since there are so many different
versions of FFTs, different parallel schemes for the FFTs, and
different parallel computers to implement them. There is an
argument that the transpose-split scheme\textit{ is no longer the
clear winner }(chapter 23 of Ref ~\cite{echu00}) because it can
not avoid the data communication while using 1D FFT in two
directions. In this paper, we will not devote our efforts to
parallel FFT (simply use the most popular transpose-split),
instead, we will try to find other methods to improve the
parallel efficiency. Our parallel scheme will be even faster if a
faster parallel FFT is adopted.

The total time ($T_{sum} $) for each processor is the sum of the
computation time ($T_{comp} $) and the communication time
($T_{comm} $):

\begin{equation}
\label{eq6} T_{sum} = T_{comp} + T_{comm} .
\end{equation}

In the case of PFFT, the 2D NS equations need to calculate ten
FFTs, which makes the major part of the computations. If the
resolution of the simulation is $M^2$, then

\begin{equation}
\label{eq7} T_{comp} = 10\times \frac{t_{FFT} }{p} =
\frac{10}{p}\times \left( {M^2\log _2 \left( {M^2} \right)\;t_c }
\right),
\end{equation}

where, $p$ is the total number of processors, $t_{FFT} $ is the
time that requires by one processor to compute one FFT, and $t_c$
is CPU time of one FFT in a single processor times a factor (the
factor is 5 in the case of full complex FFT, and 5/2 for
real-complex FFT) ~\cite{canu87}. For each FFT, one processor
needs to send ($p - 1$) blocks of data to other ($p - 1$)
processors, and receives the same amount of data from all other
processors. The size of each block is $M^2 / p^2$, so

\begin{equation}
\label{eq8} T_{comm} = 2\times \left( {(p - 1)\times
\frac{M^2}{p^2}\times t_{sendrec} + (p - 1)\times t_{delay} }
\right)\times 10,
\end{equation}

where, $t_{sendrec} $ is the time to transmit a word between
processors, and $t_{delay} $ is the latency time for a message
passing. For the convenience of the analysis later, it is useful
to divide $T_{comm} $ into two parts -- transmission time
($T_{sendrec} $) and latency time ($T_{delay} $):
\begin{eqnarray}
\label{eq9} T_{sendrec} &=& 2\times \left( {(p - 1)\times
\frac{M^2}{p^2}\times t_{sendrec} } \right)\times 10,
\\
\label{eq10} T_{delay} &=& 2\times \left( {(p - 1)\times t_{delay}
} \right)\times 10.
\end{eqnarray}
Hence, according to our analytical model, the total time for one
time step on one processor is:
\begin{eqnarray}
T_{sum} &=&  (20p - 20)\times \frac{M^2}{p^2}\times
t_{sendrec} + (20p - 20)\times t_{delay} \nn \\
&& \quad + \frac{10}{p}\times
\left( {M^2\log _2 \left( {M^2} \right)\;t_c } \right).
\label{eq11}
\end{eqnarray}
In fact, the values of $t_c $ is smaller for larger CPU numbers
because of the cache effect (see the discussion later in this
subsection), while $t_{sendrec} $ and $t_{delay} $ are almost
constant for a given parallel computer. Here, we are only trying
to give an estimated analysis and treat $t_{sendrec} $, $t_{delay}
$, and $t_c $ as constant; accurate timing work will be the
speedup plot resulting from the wall clock time (see, e.g., Fig.
1, the similar approach is also adopted in ~\cite{dmit01}).

For many parallel systems, $t_{delay} $ is much larger than
$t_{sendrec} $ (sometimes a factor of 1000 ~\cite{roos95}), so
$T_{delay} $ is a nontrivial part of $T_{sum} $, especially when
$M^2$ is small. According to Eq. (11), if we use more processors
in the simulation, $T_{delay} $ will become larger while the rest
part ($T_{sendrec} + T_{comp} $) will be smaller. Eventually,
$T_{delay} $ will become the dominating part of $T_{sum} $, which
affects the parallel efficiency of the PFFT scheme. In the case of
the 2D DNS, this phenomenon is seen in relatively small CPU
numbers, because $T_{sendrec} + T_{comp} $ is not very large. In
the 3D DNS with high resolutions ($512^3$ or higher), $T_{delay} $
will take a very small portion in $T_{sum} $ except for very large
$p$. This is the main difference between the 2D and 3D problems.

\begin{figure}[t!]
\centering
\begin{minipage}[c]{.5 \linewidth}
\scalebox{1}[1]{\includegraphics[width=\linewidth]{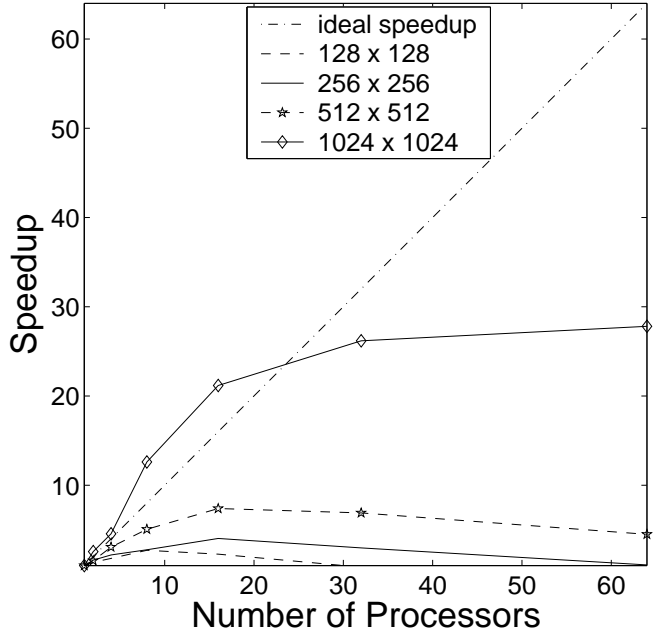}}
\end{minipage}
\caption{The speedup of 2D Navier-Stokes code with PFFT scheme
(or ``1-n'' scheme).}
\end{figure}

Fig. 1 shows the speedup of PFFT on different resolutions. The
speedup factor is the wall clock time of serial code divided by
that of the parallel code in the same resolution. For the
resolution of $128^2$, the run with eight processors has the top
speed. If the CPU number is larger than eight, the speed of the
code drops down, and the parallel code with 32 or 64 processors is
even slower than the serial code. For resolutions of $256^2$ and
$512^2$, 16 CPUs give the best performance, while 32 and 64 CPUs
lead to worse performance. For $1024^2$, the fastest run is the
one with the maximum available CPUs. (The SGI 3800 in our
laboratory is a 64-nodes system; we can not test the cases with
more than 64 processors.) We can predict from the tendency of the
curve (or, Eq. (11)) that the speedup will also drop down for the
$1024^2$ run with 128 or more processors.

One may notice that the super-linear speedup for runs with
$1024^2$ resolution: in the case of 16 CPUs, a speedup of 21 is
observed. The super-linear speedup is due to the so-called
\textit{cache effect}. In the architecture of modern computers,
caches, which can be viewed as a smaller and faster memory, are
widely used to increase the computer performance. For the PFFT NS
simulation with fixed $M^2$, the more CPUs are involved, the
higher hit rate of cache will be obtained because the code
exhibits sufficient data locality in its memory accesses. Hence,
the unit calculation finished on one CPU in the PFFT code is
faster than one CPU with the serial codes, that is, $t_c $ is
smaller. The behavior of the cache memory is very hard to predict,
because modern computer architectures have very complex internal
organizations: different levels of caches, branch predictors,
etc. It is challenging to predict the cache effect in the
research of parallel computation ~\cite{frag04,cull99}.
Therefore, we do not take the cache effect into the consideration
of the analyzing model in Eqs. (6-11).

To sum up, $T_{delay}$ takes a large portion in the total
communication time ($T_{comm} $) and seriously reduces the
parallel efficiency for larger CPU numbers in the PFFT scheme.
The issue of minimizing $T_{delay} $ is the key to enhance the
parallel efficiency.

If we start parallel programming from an available serial code,
the PTD scheme is always the first choice because it is very easy
to implement ~\cite{zyin04}. Our new parallel scheme (PTF) will
combine PTD and PFFT scheme together, with PTD being a special
case.

\subsection{A new parallel strategy - PTF}

As mentioned earlier, there are ten FFTs needed to be evaluated
at each time step. In PTF scheme, they can be divided into two,
four, and six groups corresponding to the 2, 4, and 6 CPUs scheme
in PTD ~\cite{zyin04}. In the following, we will call these three
PTF schemes as ``2-n,'' ``4-n,'' and ``6-n'' scheme,
respectively. (We also denote PFFT scheme as ``1-n'' scheme to
unify notations). There are six FFTs in the 2-n scheme for each
group, while three FFTs in the 4-n scheme. Note that there are
twelve FFTs in the 2-n and 4-n schemes at each time step (there
is only ten FFTs in serial and PFFT codes) because the two FFTs
in the 2$^{nd}$ row of Table 1 are calculated twice to eliminate
the total communication time.

The 6-n scheme will not be discussed in the rest of the paper,
because we want to compare the parallel efficiency of the PTF
scheme with the PFFT (1-n) scheme, the number of CPUs involved
should be powers of 2.

For any type of PTF scheme, one group will be called ``master
group,'' on which the time integration and data input and output
(I/O) are carried out, and the other groups will be called ``slave
groups.'' When calculating FFT, data are exchanged only within
each group. When it is necessary to transfer information between
groups, the first node in master group will and only will
communicate with the first nodes in the slave groups; likewise,
the second nodes in different groups will communicate with each
other, etc.

For the 2-n scheme, the computation time of one processor is

\begin{equation}
\label{eq12} T_{comp} = 6\times \frac{t_{FFT} }{p / 2} =
\frac{12}{p}\times \left( {M^2\log _2 \left( {M^2} \right)\;t_c }
\right),
\end{equation}

where, the factor 6 comes from the six FFTs in each group, and
$t_{FFT} $ is divided by $p / 2$ because the processors available
are split into two groups. In the beginning of time loop, each
node in the master group needs to send one $\frac{M^2}{p / 2}$
block data to the corresponding node in the slave group, and
receive the same amount of data at the end of time loop. So
together with the data transferred within each FFT, the two parts
of the communication time are:
\begin{eqnarray}
\label{eq13} T_{sendrec} &=& 2\times \left( {(\frac{p}{2} - 1)\times
\frac{M^2}{\left( {p / 2} \right)^2}\times 6 + \frac{M^2}{p / 2}}
\right)\times t_{sendrec} = \left( {28p - 4\mbox{8}}
\right)\times \frac{M^2t_{sendrec} }{p^2}
\\
\label{eq14} T_{delay} &=& \left( {2\times \left( {\frac{p}{2} - 1}
\right)\times 6 + 2} \right)\times t_{delay} = (6p - 10)\times
t_{delay} .
\end{eqnarray}
The total time per time step is
\begin{equation}
\label{eq15} T_{sum} = (2\mbox{8}p - 4\mbox{8})\times
\frac{M^2}{p^2}\times t_{sendrec} + (6p - 10)\times t_{delay} +
\frac{12}{p}\times \left( {M^2\log _2 \left( {M^2} \right)\;t_c }
\right).
\end{equation}
Similarly, for the 4-n scheme, there is one master group and three
slave groups. The computation time is
\begin{equation}
\label{eq16} T_{comp} = 3\times \frac{t_{FFT} }{p / 4} =
\frac{12}{p}\times \left( {M^2\log _2 \left( {M^2} \right)\;t_c }
\right).
\end{equation}
In the beginning of time loop, each node in the master group needs
to broadcast one $\frac{M^2}{p / 4}$ block data to the
corresponding node in slave groups, and gather the same amount of
data at the end of time loop. The standard broadcast and gather
routine in MPI require $\log_2^g$ steps ($g$ is the size of the
group; here, $g = 4$) in sending and receiving to finish the
group communication. So together with the data transferred within
each FFT, the two parts of the communication time are:
\begin{eqnarray}
T_{sendrec} &=&  2\times \left( {(\frac{p}{4} -
1)\times \frac{M^2}{\left( {p / 4} \right)^2}\times 3 +
\frac{M^2}{p / 4}\times \log _2^4 } \right)\times t_{sendrec} \nn \\
&=& \left( {40p - 96} \right)\times \frac{M^2t_{sendrec} }{p^2},
\label{eq17}\\
T_{delay} &=& \left( {2\times \left( {\frac{p}{4} - 1}
\right)\times 3 + 2\times \log _2^4 } \right)\times t_{delay} =
(\frac{3p}{2} - 2)\times t_{delay} ,
\label{eq18}
\end{eqnarray}
and the total time used per time step is
\begin{equation}
\label{eq19} T_{sum} = (40p - 96)\times \frac{M^2}{p^2}\times
t_{sendrec} + (\frac{3p}{2} - 2)\times t_{delay} +
\frac{12}{p}\times \left( {M^2\log _2 \left( {M^2} \right)\;t_c }
\right).
\end{equation}

As can be seen from Eqs. (11), (15), and (19), $T_{comp} $ in the
1-n, 2-n, and 4-n schemes are roughly the same (20{\%} difference
at most) for fixed $M$ and $p$; $T_{delay} $ in the 1-n scheme is
about three times as large as that in the 2-n scheme, and about
thirteen times as large as that in 4-n scheme. In the meanwhile,
$T_{sendrec} $ in the 2-n or 4-n scheme is increased by
relatively small factor (no more than 2) compared with 1-n
scheme. Hence, $T_{delay} $ in the 1-n scheme will occupy the
largest partition of $T_{sum} $ among the 1-n, 2-n, and 4-n
schemes.

As discussed in the previous subsection, $T_{delay} $ is the main
barrier for achieving high parallel efficiency when $p$ is large
for relatively low resolution ($M^2$). This problem is partly
solved by using 2-n and 4-n schemes. However, even in the 2-n and
4-n schemes, $T_{delay}$ still increases and the remaining parts
of $T_{sum} $ decrease for larger $p$; the bottleneck effect in
parallelization still exits.

\begin{figure}[t!]
\centering
\begin{minipage}[c]{.4 \linewidth}
\scalebox{1}[1]{\includegraphics[width=\linewidth]{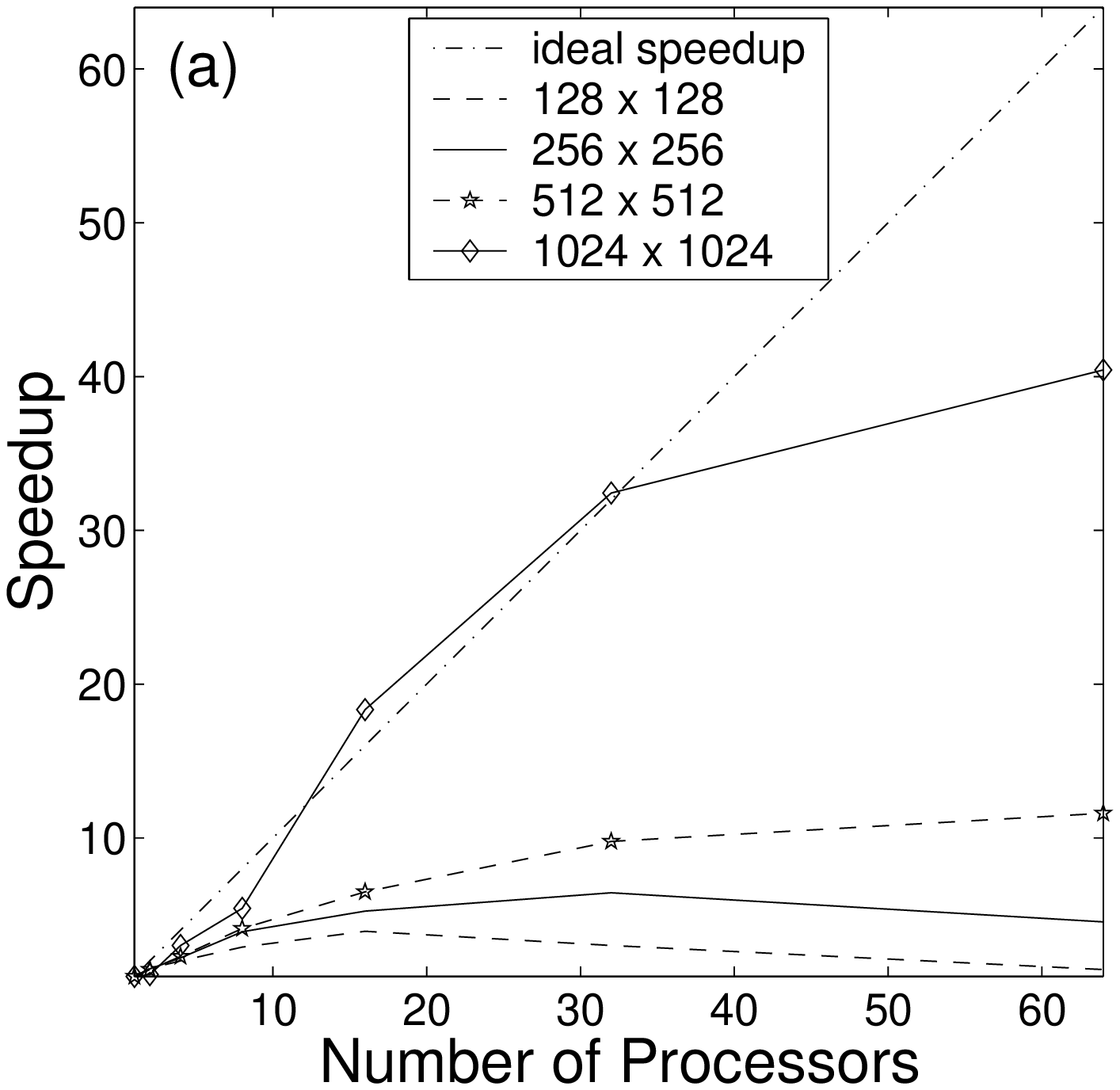}}
\end{minipage}
\begin{minipage}[c]{.4 \linewidth}
\scalebox{1}[1]{\includegraphics[width=\linewidth]{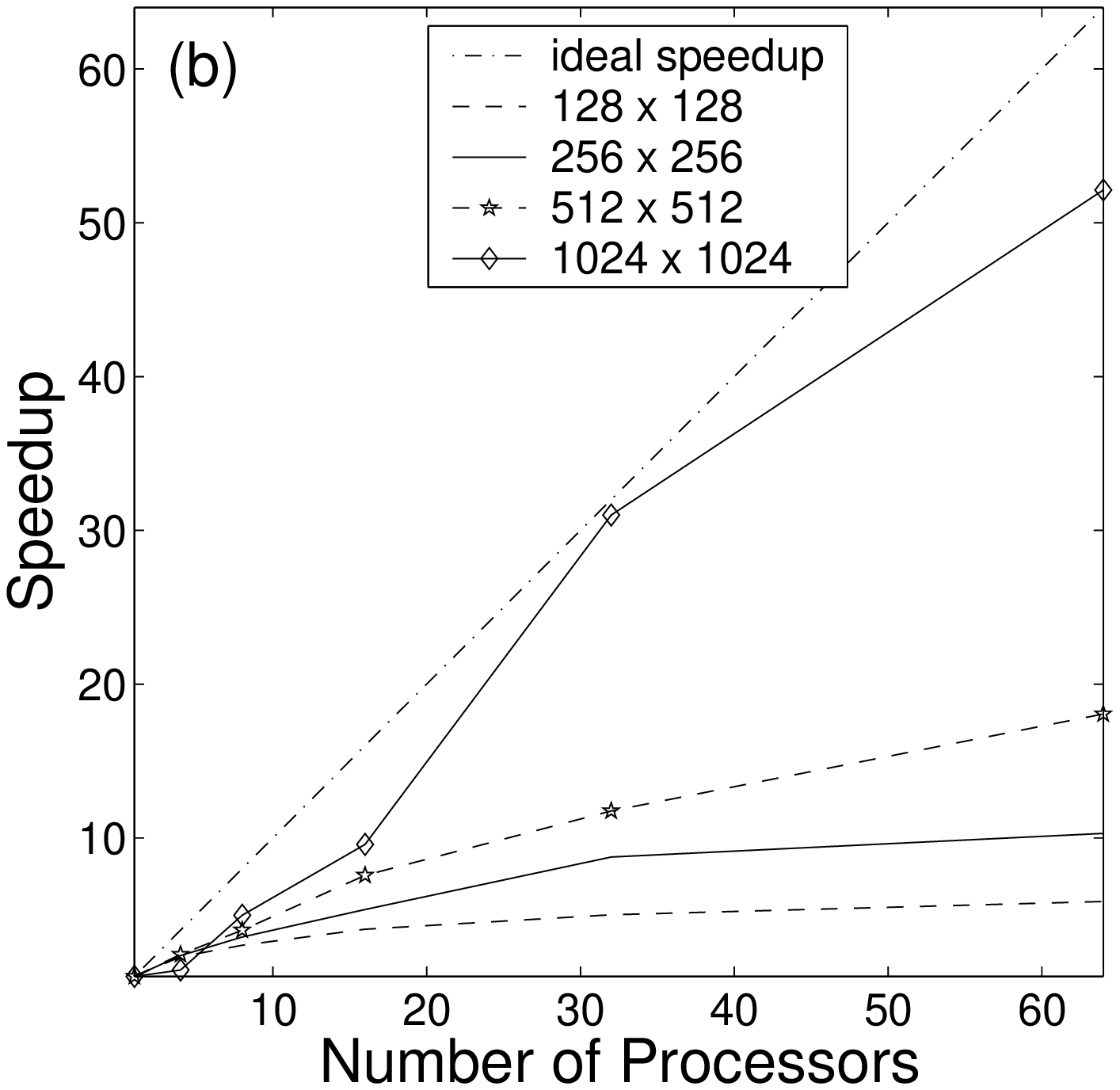}}
\end{minipage}
\caption{The speedup of 2D Navier-Stokes code with ``2-n'' scheme
(a) and ``4-n'' scheme (b).}
\end{figure}

Figs. 2(a) shows the speedup plot for the 2-n scheme. For the
resolution of $128^2$, the peak speed appears at 16 CPUs run
(compared with 8 CPUs run for the 1-n scheme). For the resolution
of $256^2$, the peak speed appears at 32 CPUs run (compared with
16 CPUs run for 1-n scheme). For $512^2$ and $1024^2$ in the 2-n
scheme, and all the resolutions in the 4-n scheme (Fig. 2(b)), 64
CPUs run is the fastest.

Because two extra FFTs are introduced and $T_{sendrec}$ is
slightly larger in 2-n and 4-n schemes, some runs in Figs. 2 are
slower than the corresponding 1-n ones for certain resolution and
$p$. For example, there is no super-linear speedup observed for
the $1024^2$ curve in Fig. 2(b), although the 4-n scheme is twice
as fast as the 1-n scheme when 64 processors are used.

In practice, the fastest scheme for fixed $p$ and $M^2$ is always
what we want to use to do long time simulation. Fig.3 is a
combined figure which consists of the best performance point in
Figs. 1 and 2. Table 2 indicates the corresponding schemes
adopted to draw this ``best speedup'' plot.

It should be emphasized that most points in the $1024^2$ curve
show super-linear speedup. The points when $p \le 16$ come from
the 1-n scheme, while the 2-n and 4-n schemes are faster for $p
\ge 32$.

\begin{figure}[t!]
\centering
\begin{minipage}[c]{.5 \linewidth}
\scalebox{1.1}[1]{\includegraphics[width=\linewidth]{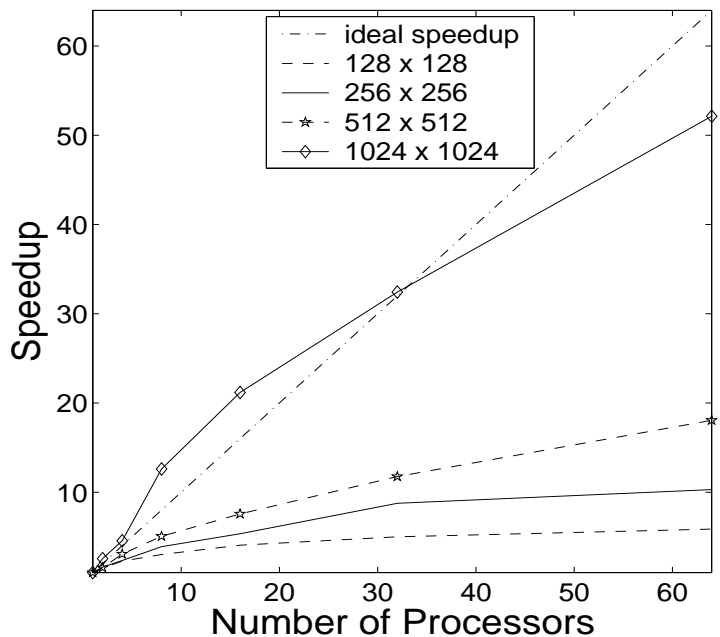}}
\end{minipage}
\caption{The speedup of 2D Navier-Stokes code with ``ideal''
scheme. The values used here are the top speedup among 1-n, 2-n,
and 4-n scheme for same resolutions and same numbers of
processors.}
\end{figure}

\begin{table}[t!]
\caption{The corresponding schemes for the values in Fig. 3.}
\centering
\begin{tabular}
{|p{54pt}|p{34pt}|p{34pt}|p{34pt}|p{34pt}|} \hline \hline
\textbf{}& $128^2$\textbf{}& $256^2$\textbf{}& $512^2$\textbf{}&
$1024^2$\textbf{} \\
\hline 1 CPU& 1-n& 1-n& 1-n&
1-n \\
\hline 2 CPUs& 2-n& 1-n& 1-n&
1-n \\
\hline 4 CPUs& 4-n& 4-n& 1-n&
1-n \\
\hline 8 CPUs& 4-n& 2-n& 1-n&
1-n \\
\hline 16 CPUs& 4-n& 4-n& 4-n&
1-n \\
\hline 32 CPUs& 4-n& 4-n& 4-n&
2-n \\
\hline 64 CPUs& 4-n& 4-n& 4-n&
4-n \\
\hline \hline
\end{tabular}
\end{table}

For lower resolutions, the 1-n scheme works the best for small
$p$, and the 2-n or 4-n scheme dominates the points on the
corresponding curve in Fig.3 gradually for larger $p$. On the
first column of Table 2, which corresponds to the resolution of
$128^2$, the 4-n scheme dominates for $p \ge 4$.

The situations of $p = 2$ in the 2-n scheme and $p = 4$ in the
4-n scheme correspond to the PTD scheme. Although the benchmark
results are slightly different from ~\cite{zyin04} due to
different computers used, the conclusion is the same: PTD is an
easily implemented and efficient parallel scheme for the 2D DNS
for relatively small resolutions.

The easy implementary property of PTD is inherited by PTF
schemes: once the PFFT code is ready, it is very easy to change
it to PTF codes. Hence, PTF is an attractive strategy, especially
in 2D simulations.

\subsection{Numerical results for 2D decaying turbulence}

In this subsection, we will use our PTF codes to investigate an
interesting phenomenon in 2D decaying turbulence -- the
multi-valued $\omega$-$\psi $ structure.

It is well known that if there is a functional relation $\omega =
f(\psi )$, then the nonlinear term in Eq. (3) gives:

\begin{equation}
\label{eq20}
 {\rm {\bf u}} \cdot \nabla \omega = \frac{\partial \psi }{\partial
y}\frac{\partial \omega }{\partial x} - \frac{\partial \psi
}{\partial x}\frac{\partial \omega }{\partial y} = \frac{\partial
f(\psi )}{\partial x}\frac{\partial \psi }{\partial y} -
\frac{\partial f(\psi )}{\partial y}\frac{\partial \psi
}{\partial x}= (\frac{\partial f}{\partial \psi }\frac{\partial
\psi }{\partial x})\frac{\partial \psi }{\partial y} -
(\frac{\partial f}{\partial \psi }\frac{\partial \psi }{\partial
y})\frac{\partial \psi }{\partial x} = 0.
\end{equation}

Thus Eq. (3) becomes

\begin{equation}
\label{eq21} \frac{\partial \omega }{\partial t} = \nu \Delta
\omega.
\end{equation}

For very high Reynolds number or $\nu \to 0$, Eq. (21) turns to a
stationary equation

\begin{equation}
\label{eq22} \frac{\partial \omega }{\partial t} = 0.
\end{equation}

It is common to treat $\omega = f(\psi )$ as an indication of the
final state of 2D turbulence, but the simulation shown in Figs.
18-19 of Ref ~\cite{zyin03} gives a counter example: the
double-valued $\omega$-$\psi $ structure (see ~\cite{zyin04a} for
a detailed discussion about this kind of structure). We will seek
some other simulations to validate the generality of multi-valued
structures, and there are mainly two choices to do this:

\begin{itemize}
\item Make changes in the initial conditions;
\item Test different values of $\nu$ for certain kind of initial
condition.
\end{itemize}

The parallel code we used before (PTD scheme ~\cite{zyin04}) has
a very limited speedup because the maximum number of CPUs used is
limited to six. Most simulations were carried out by changing the
initial condition for relatively small $\nu$ with the resolution
of $512^2$ ~\cite{zyin04a}, which normally last from several days
to a few weeks. For $1024^2$ runs, the calculations for one time
step are four times as large as those for $512^2$, and the time
step has to be smaller due to the CFL condition. A typical 2D
decaying turbulence lasts from two months to half year if we only
use PTD schemes. With the PTF codes, it is now possible to carry
out $1024^2$ simulations to find the new double-valued $\omega$-$
\psi$ structure.

We carried out two simulations starting from four equal sized
vortices patches, which are asymmetrically placed in a double
periodic box (Fig. 4). The first run adopted relatively low
Reynolds number ($1 / \nu = 4000$) with the resolution of $512^2$.
The time step is 0.0005. We used 32 CPUs in total (this is the
maximum nodes available for long time simulations in LSEC).
According to Table 2, the 4-n scheme is the fastest approach under
this situation. The simulation lasted for 17 hours, and reached
the quasi-steady state at $t \approx 200$ (Fig. 5(a)). Fig. 5(b)
shows a simple function relation of $\omega$-$\psi$. The same run
lasts for 29 hours if we use the PFFT scheme (32 CPUs), and 82
hours for the PTD scheme (4 CPUs).

\begin{figure}[t!]
\centering
\begin{minipage}[c]{.5 \linewidth}
\scalebox{1}[1]{\includegraphics[width=\linewidth]{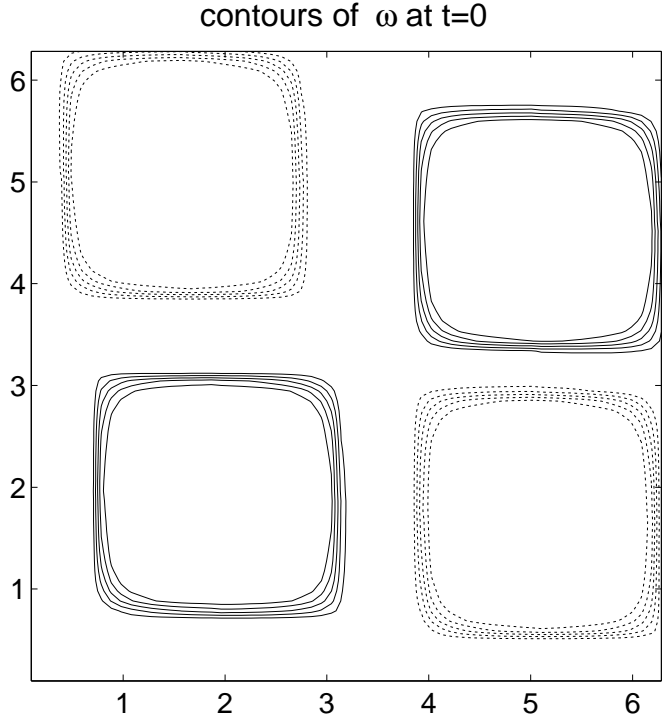}}
\end{minipage}
\caption{Contour plot of the initial condition in a simulation of
Navier-Stoke equation which is initialized by two positive and
two negative flat vortices (these four vortices are placed
asymmetrically in the flow field). Dashed contours represent
positive vorticity and drawn contours represent negative
vorticity. }
\end{figure}

\begin{figure}[!htbp]
\centering
\begin{minipage}[c]{.4 \linewidth}
\scalebox{1}[1]{\includegraphics[width=\linewidth]{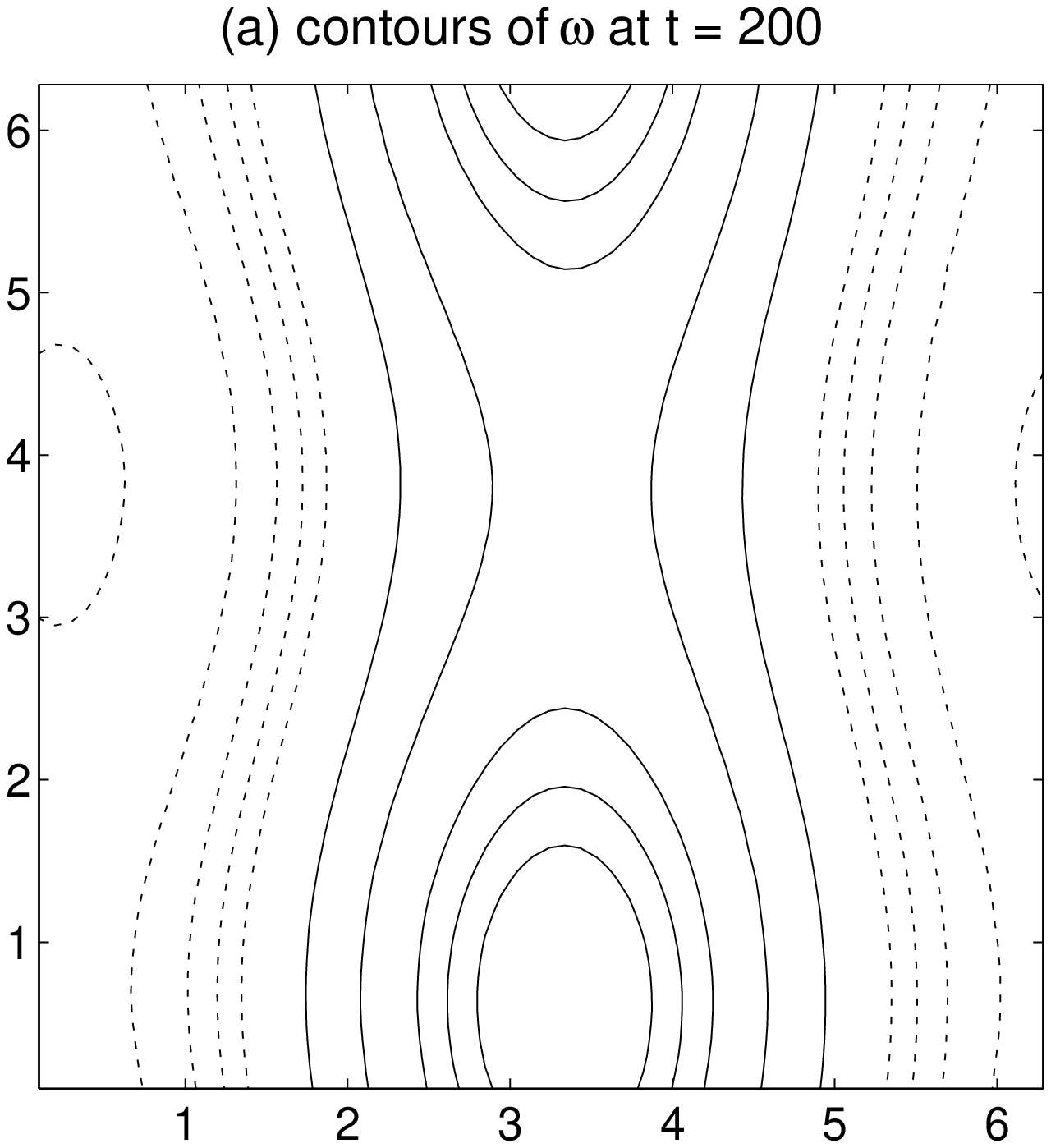}}
\end{minipage}
\begin{minipage}[c]{.4 \linewidth}
\scalebox{1}[1.2]{\includegraphics[width=\linewidth]{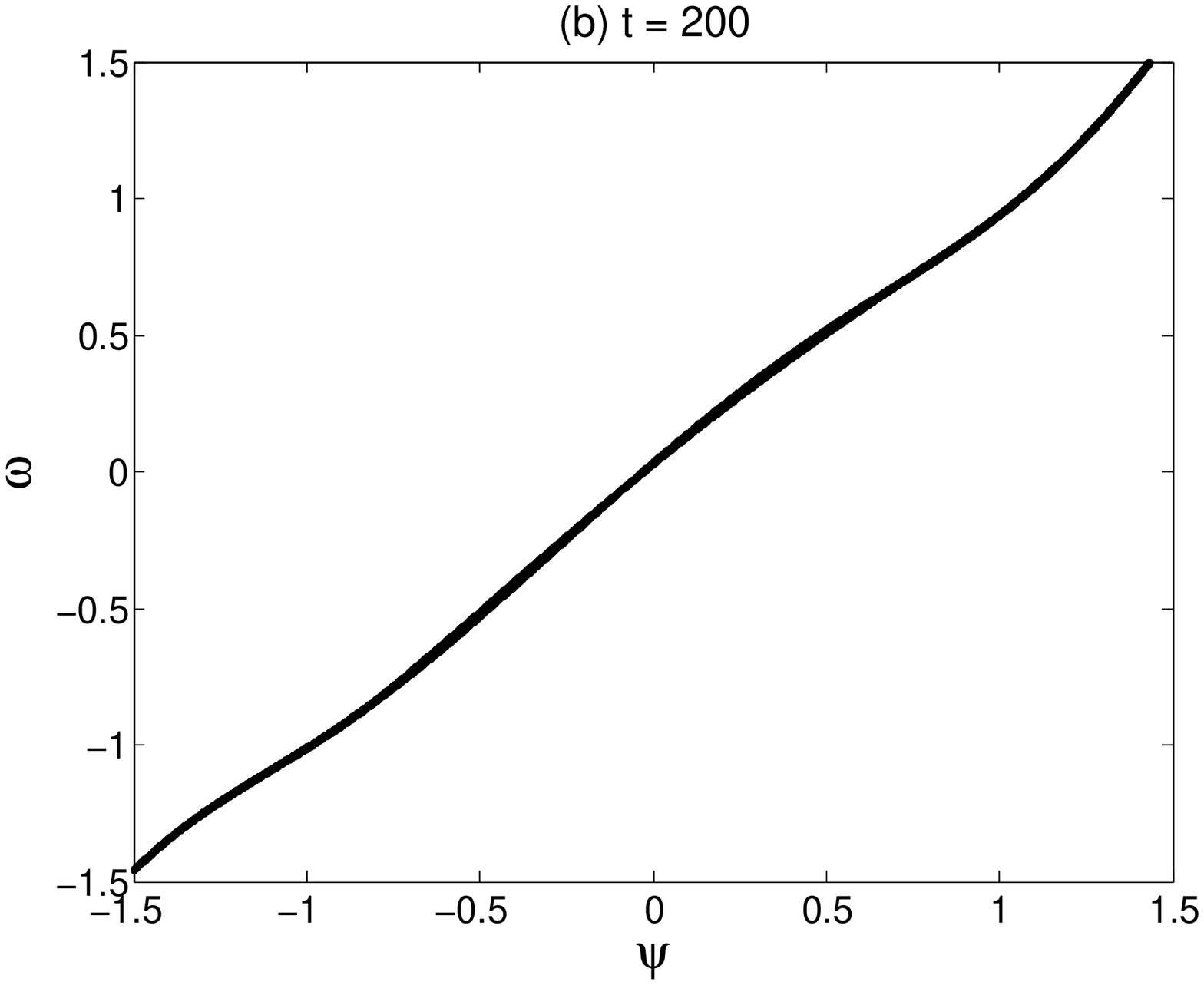}}
\end{minipage}
\caption{(a) is the vorticity contour plot at $t = 200$ evolving
from the initial condition shown in Fig. 4 (Re=4000). The
resolution adopted is $512^2$. (b) is the $\omega$-$\psi $
scatter plot at the same time, which shows a simple functional
relation.}
\end{figure}

\begin{figure}[!htbp]
\centering
\begin{minipage}[c]{.4 \linewidth}
\scalebox{1}[1]{\includegraphics[width=\linewidth]{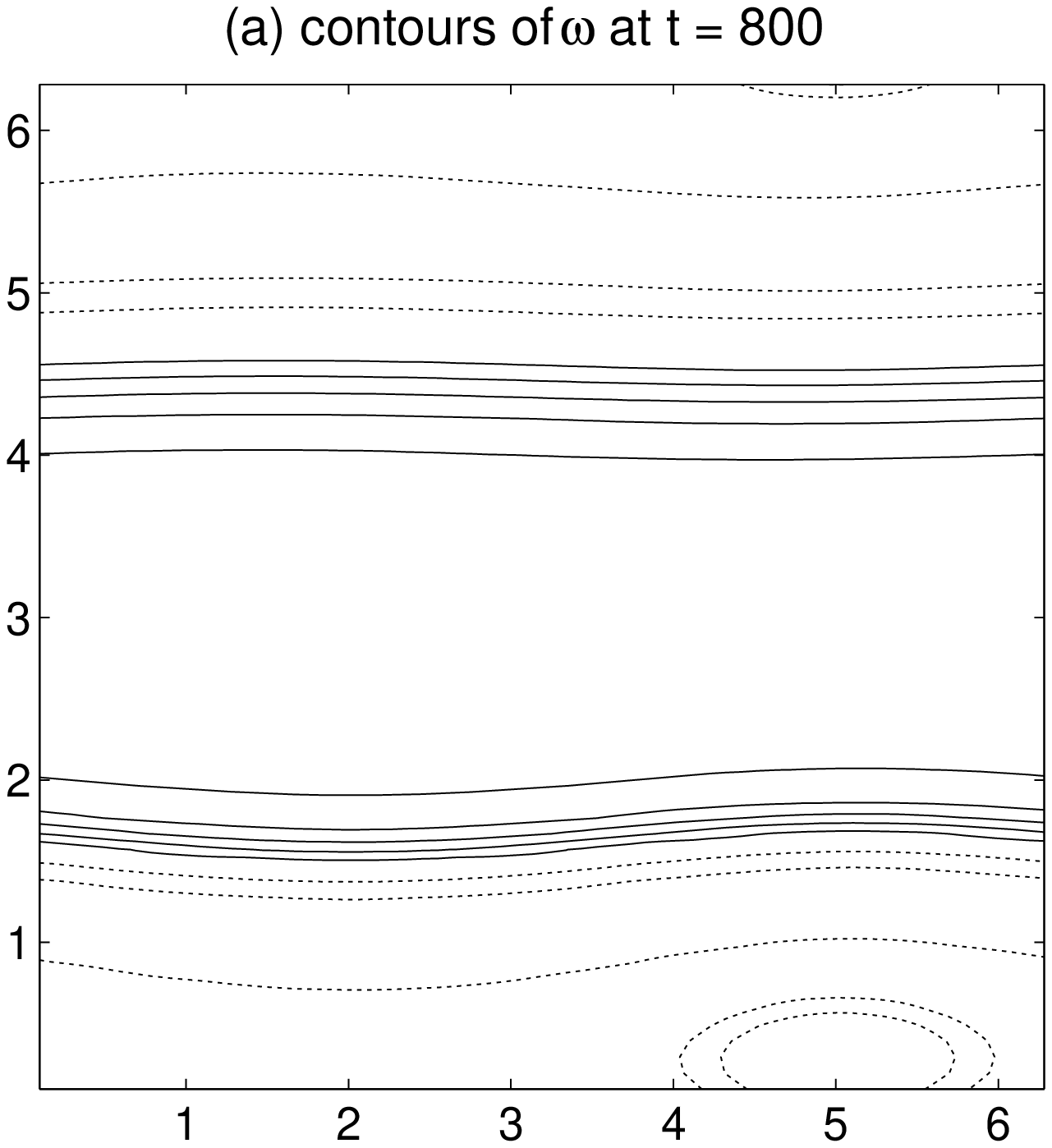}}
\end{minipage}
\begin{minipage}[c]{.4 \linewidth}
\scalebox{1}[1.2]{\includegraphics[width=\linewidth]{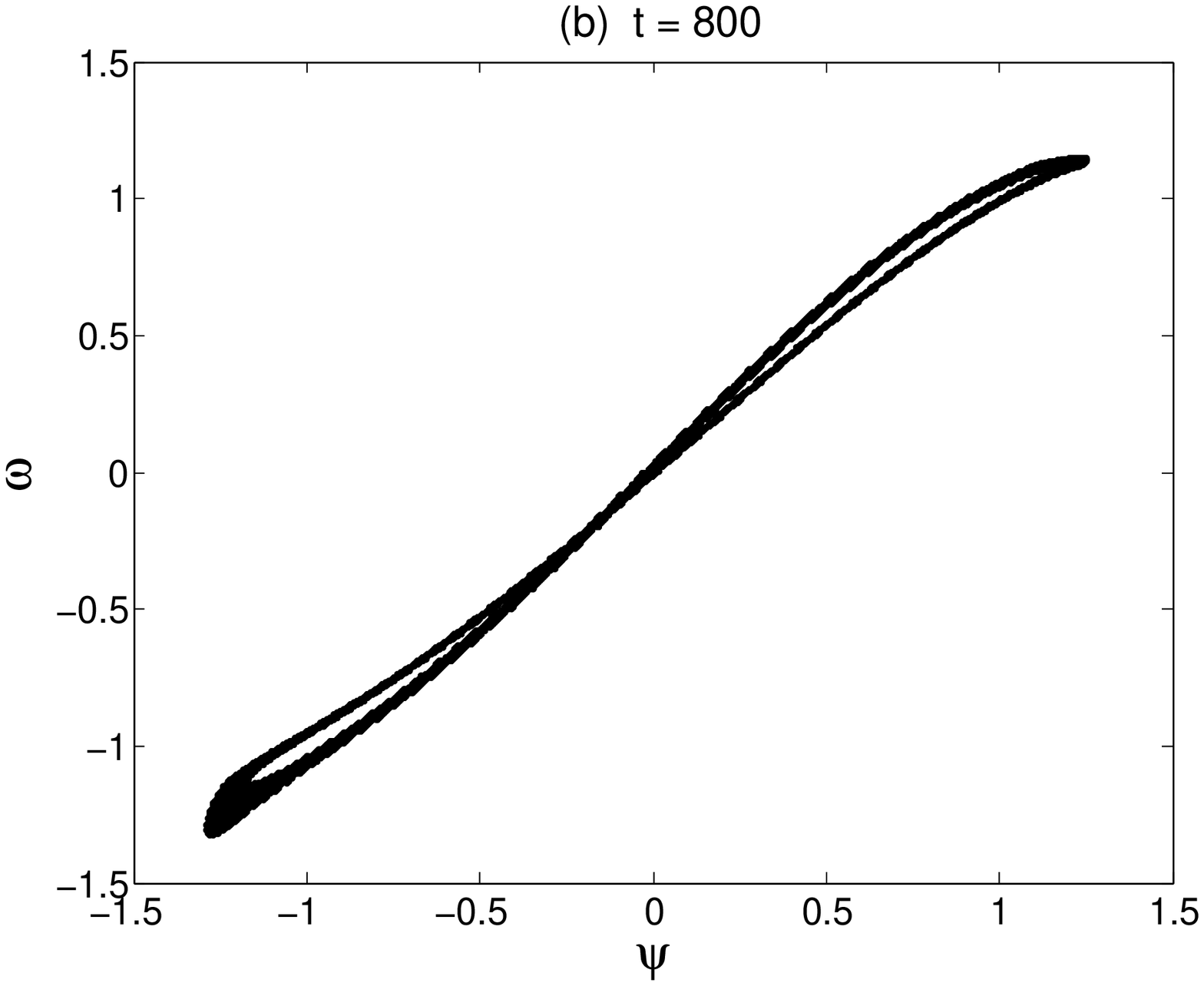}}
\end{minipage}
\caption{(a) is the vorticity contour plot at $t = 800$ evolving
from the initial condition shown in Fig. 4 (Re=40000). The
resolution adopted is $1024^2$. (b) is the $\omega$-$\psi $
scatter plot at the same time, which shows a double-valued
structure.}
\end{figure}

The second run adopted a large Reynolds number ($1 / \nu = 40000$)
with the resolution of $1024^2$. We used the 2-n scheme code on
32 processors, and the quasi-steady state was reached at $t
\approx 800$ after 16 days' calculation. In the late state of
this simulation, the orientation of the flow pattern is totally
different from that obtained with the $512^2$ run (Fig. 6(a)).
Moreover, the double-valued $\omega$-$\psi $ structure is
recovered (Fig. 6(b)), which fits the definition of the final
state of the 2D turbulence given in ~\cite{zyin04a}. Hence, it
seems that further computations do not lead to any new
phenomenon. The same run lasts for 20 days if we use the PFFT
scheme (32 nodes), and about one year for the PTD scheme (4
nodes).

It is interesting to see the appearance of multi-valued $\omega$-$
\psi $ structures in high Reynolds number (40000), while
relatively low Reynolds number (4000) leads to traditional
functional relation. Although the physical mechanism of the
multi-valued structure is still unclear, our simulations reveal
some limitations of the statistical mechanics~\cite{zyin04a}, and
it is worthwhile to carry on further investigations in this
direction. Of course, the PTF scheme may play an important role
in these large scale computations because of its high parallel
efficiency.

With maximum 32 processors available for long time simulation,
the peak speed of PFT is only about 70{\%} faster than that of
PFFT for the resolution of $512^2$, and 24{\%} for $1024^2$.
However, for runs like what are shown in Figs. 6, 24{\%} shortage
still means 4 days' calculation on 32 nodes, which is definitely
nontrivial.

\section{Application of PTF scheme to 2D inviscid Boussinesq equations}

It is very interesting to understand whether a finite time
blow-up of the vorticity and temperature gradient can happen from
a smooth initial condition in 2D inviscid Boussinesq convection.
This has been studied recently by several groups ending with
different conclusions: Pumir and Siggia used the TVD scheme with
the adaptive meshes (the maximum grid size is $256^2$), which
observed the blow-up process ~\cite{pumi92}; E and Shu did not
obtain the blow-up phenomena using the ENO scheme on a $512^2$
grid and spectral methods on a $1500^2$ grid ~\cite{we94}; and
Ceniceros and Hou also obtained blow-up free solutions using the
adaptive mesh computation with a $512^2$ grid~\cite{ceni01}. It is
worth mentioning that the maximum mesh compression ratio
in~\cite{ceni01} is 8.83, which gave an effective resolution of
$4600^2$ on uniform mesh. Although those groups yielded different
conclusions, they all tried to use the highest possible
resolutions to investigate the problem. Hence, parallel computing
is a natural choice.

The 2D inviscid Boussinesq convection equations can be written in
$\omega$-$\psi $ formulation:
\begin{eqnarray}
&& \rho _t + {\rm {\bf u}} \cdot \nabla \rho = 0,
\label{eq23}
\\
&& \omega _t + {\rm {\bf u}} \cdot \nabla \omega = -
\rho _x,
\label{eq24}
\\
&& \Delta \psi = - \omega.
\label{eq25}
\end{eqnarray}
Again, ABCN scheme is adopted to carry out the time integration.
Below we will discuss how to implement our parallel strategy for
Eqs. (23) - (25). Numerical results on the finite time blow-up
will be reported.

\subsection{The application of the new strategy to Boussinesq
equations}

As shown in Table 3, there are 20 FFTs involved when the
Fourier-Galerkin spectral methods are used to solve Eqs.
(23)-(25). These FFTs can be divided into four independent
groups, which are indicated by the different columns in Table 3.
There is no communication within the different columns until all
FFTs are finished. In each column, the 1$^{st }$ and the 2$^{nd}$
FFT need to be evaluated before the 4$^{th}$ one, while the
5$^{th}$ FFT should be calculated after the 1$^{st}$ and 3$^{rd}$
one are finished.

\begin{table}[t!]
\caption{Twenty FFTs need to be calculated in each time step of
Boussinesq spectral solver.} \centering
\begin{tabular}
{|p{17pt}|p{78pt}|p{78pt}|p{78pt}|p{78pt}|} \hline \hline & A& B&
C&
D \\
\hline 1& $\hat{\mbox{u}} \to \mbox{u}$& $\hat{\mbox{U}} \to
\mbox{U}$& $\hat{\mbox{v}} \to \mbox{v}$&
$\hat{\mbox{V}} \to \mbox{V}$ \\
\hline 2& $\hat {\omega }_x \to \omega _x $& $\hat {\Omega }_x
\to \Omega _x $& $\hat {\omega }_y \to \omega _y $&
$\hat {\Omega }_y \to \Omega _y $ \\
\hline 3& $\hat {\rho }_x \to \rho _x $& $\hat {P}_x \to P_x $&
$\hat {\rho }_y \to \rho _y $&
$\hat {P}_y \to P_y $ \\
\hline 4& $\mbox{u}\omega _x \to \widehat{\left( {\mbox{u}\omega
_x } \right)}$& $\mbox{U}\Omega _x \to \widehat{\left(
{\mbox{U}\Omega _x } \right)}$& $\mbox{v}\omega _y \to
\widehat{\left( {\mbox{v}\omega _y } \right)}$&
$\mbox{V}\Omega _y \to \widehat{\left( {\mbox{V}\Omega _y } \right)}$ \\
\hline 5& $\mbox{u}\rho _x \to \widehat{\left( {\mbox{u}\rho _x }
\right)}$& $\mbox{UP}_x \to \widehat{\left( {\mbox{UP}_x }
\right)}$& $\mbox{v}\rho _y \to \widehat{\left( {\mbox{v}\rho _y
} \right)}$&
$\mbox{VP}_y \to \widehat{\left( {\mbox{VP}_y } \right)}$ \\
\hline \hline
\end{tabular}
\end{table}
\begin{table}[t!]
\caption{In 8-N scheme, the total twenty four FFTs in each time
step of Boussinesq spectral solver are divided into eight groups
(A-H). There are four FFTs introduced in addition to the twenty
FFTs in Table 3.} \centering
\begin{tabular}
{|p{22pt}|p{78pt}|p{78pt}|p{78pt}|p{78pt}|} \hline\hline & A& B&
C&
D \\
\hline 1& $\hat{\mbox {u}} \to \mbox{u}$& $\hat{\mbox {U}} \to
\mbox{U}$& $\hat{\mbox {v}} \to \mbox{v}$&
$\hat{\mbox {V}} \to \mbox{V}$ \\
\hline 2& $\hat {\omega }_x \to \omega _x $& $\hat {\Omega }_x
\to \Omega _x $& $\hat {\omega }_y \to \omega _y $&
$\hat {\Omega }_y \to \Omega _y $ \\
\hline 3& $\mbox{u}\omega _x \to \widehat{\left( {\mbox{u}\omega
_x } \right)}$& $\mbox{U}\Omega _x \to \widehat{\left(
{\mbox{U}\Omega _x } \right)}$& $\mbox{v}\omega _y \to
\widehat{\left( {\mbox{v}\omega _y } \right)}$&
$\mbox{V}\Omega _y \to \widehat{\left( {\mbox{V}\Omega _y } \right)}$ \\
\hline \hline & E& F& G&
H \\
\hline 1& $\hat{\mbox{u}} \to \mbox{u}$& $\hat{\mbox{U}} \to
\mbox{U}$& $\hat{\mbox{v}} \to \mbox{v}$&
$\hat{\mbox{V}} \to \mbox{V}$ \\
\hline 2& $\hat {\rho }_x \to \rho _x $& $\hat {P}_x \to P_x $&
$\hat {\rho }_y \to \rho _y $&
$\hat {P}_y \to P_y $ \\
\hline 3& $\mbox{u}\rho _x \to \widehat{\left( {\mbox{u}\rho _x }
\right)}$& $\mbox{UP}_x \to \widehat{\left( {\mbox{UP}_x }
\right)}$& $\mbox{v}\rho _y \to \widehat{\left( {\mbox{v}\rho _y
} \right)}$&
$\mbox{VP}_y \to \widehat{\left( {\mbox{VP}_y } \right)}$ \\
\hline \hline
\end{tabular}
\end{table}

When the PTF scheme is used to parallelize the code, there are
five options to divide the groups without too much extra data
communication:
\begin{enumerate}
\item 1 - N scheme -- one group; each FFT is computed by all the CPUs
involved, i.e. PFFT scheme. (Here we use ``N'' instead of ``n''
to avoid conflicts with the parallel schemes for the NS
equations).
\item 2 - N scheme -- two groups; column A {\&} B in one group, and
column C {\&} D in the other group.
\item 4 - N scheme -- four groups, which belongs to different columns
in Table 3, respectively.
\item 8 - N scheme -- eight groups, see column A-H of Table 4. Note
that there are four extra FFTs introduced to save the
communication time:. The first row FFTs in Table 3 are calculated
twice in the first row of Table 4.
\item 12 - N scheme -- twelve groups; the first three rows FFTs in
Table 3 are calculated simultaneously in twelve groups, the
results from the four groups calculating the first row FFTs are
transferred to the eight groups computing the 2$^{nd}$ and
3$^{rd}$ row respectively, and the rest eight FFTs in the
4$^{th}$ and 5$^{th}$ rows are performed in those eight groups.
(Like the 6-n scheme in the NS solver, the 12-N scheme will not be
discussed here because the number of processors involved is not of
the power of 2.)
\end{enumerate}

Similar to Section 2, we will analyze the total computation time
for different schemes in the following:

\begin{itemize}
\item  1-N scheme, 20 FFTs are involved:
\begin{equation}
\label{eq26} \textstyle
\begin{array}{l}
 T_{sum} = 20 \times (p - 1) \times 2 \times \frac{\textstyle M^2}{\textstyle p^2} \times t_{sendrec}
+ 20\times (p - 1)\times 2\times t_{delay} \\
 \quad \,\,\,\, +\frac{\textstyle 20}{\textstyle p} \times \left(
{M^2\log _2 \left( {M^2} \right)\;t_c } \right) \\
 \quad \,\,\,\, = (40p - 40) \times \frac{\textstyle M^2}{\textstyle p^2} \times t_{sendrec} + (40p
- 40) \times t_{delay} + \frac{\textstyle 20}{\textstyle p}
\times \left( {M^2\log _2 \left( {M^2}
\right)\;t_c } \right). \\
 \end{array}
\end{equation}
\item  2-N scheme, 10 FFTs in each groups, and two blocks (there are
two equations in the system - Eqs. (23) and (24)) of
$\frac{\textstyle M^2}{\textstyle p / 2}$ size data will be
transferred between the master group and the slave group in the
beginning and end parts of the time loop:
\begin{equation}
\label{eq27}
\begin{array}{l}
 T_{sum} = \left[ {10\times (\frac{\textstyle  p}{\textstyle 2} - 1)\times 2\times
\frac{\textstyle M^2}{\textstyle \left( {p / 2} \right)^2} +
2\times 2\times \frac{\textstyle M^2}{\textstyle p / 2}}
\right]\times t_{sendrec} \\
 \quad \quad \quad + \left( {10\times (\frac{\textstyle p}{\textstyle 2} - 1)\times 2 + 2\times 2}
\right)\times t_{delay} + \frac{\textstyle 10}{\textstyle p /
2}\times \left( {M^2\log _2 \left({M^2} \right)\;t_c } \right) \\
 \;\quad = (48p - 80)\times \frac{\textstyle M^2}{\textstyle p^2}\times t_{sendrec} + (10p -
16)\times t_{delay} + \frac{\textstyle 20}{\textstyle p} \times
\left( {M^2\log _2 \left( {M^2}
\right)\;t_c } \right). \\
 \end{array}
\end{equation}
\item 4-N scheme, 5 FFTs in each groups, and two blocks of
$\frac{\textstyle M^2}{\textstyle p / 4}$ size data will be
transferred between the master group and the slave groups in the
beginning and end parts of the time loop:
\begin{equation}
\label{eq28}
\begin{array}{l}
 T_{sum} = \left[ {5\times (\frac{\textstyle p}{\textstyle 4} - 1)
 \times 2\times \frac{\textstyle M^2}{\textstyle \left(
{p / 4} \right)^2} + 2\times 2\times \log _2^4 \times
\frac{\textstyle M^2}{\textstyle p / 4}}
\right]\times t_{sendrec} \\
 \quad \quad \quad + \left( {5\times (\frac{\textstyle p}{\textstyle 4} - 1)\times 2 + 2\times
2\times \log _2^4 } \right)\times t_{delay} + \frac{\textstyle
5}{\textstyle p / 4}\times \left(
{M^2\log _2 \left( {M^2} \right)\;t_c } \right) \\
 \quad \quad = (72p - 160)\times \frac{\textstyle M^2}{\textstyle p^2}\times t_{sendrec} +
(\frac{\textstyle 5}{\textstyle 2}p - 2)\times t_{delay} +
\frac{\textstyle 20}{\textstyle p}\times \left( {M^2\log _2
\left( {M^2} \right)\;t_c } \right). \\
 \end{array}
\end{equation}
\item 8-N scheme, 3 FFTs in each group, and two blocks of
$\frac{\textstyle M^2}{\textstyle p / 8}$ size data will be
transferred between the master group and the slave groups in the
beginning and end parts of the time loop:
\begin{equation}
\label{eq29}
\begin{array}{l}
 T_{sum} = \left[ {3\times (\frac{\textstyle p}{\textstyle 8} - 1)
 \times 2\times \frac{\textstyle M^2}{\textstyle \left(
{p / 8} \right)^2} + 2\times 2\times \log _2^8 \times
\frac{\textstyle M^2}{\textstyle p / 8}}
\right]\times t_{sendrec} \\
 \quad \quad \quad + \left( {3\times (\frac{\textstyle p}{\textstyle 8} - 1)\times 2 + 2\times
2\times \log _2^8 } \right)\times t_{delay} + \frac{\textstyle
3}{\textstyle p / 8}\times \left(
{M^2\log _2 \left( {M^2} \right)\;t_c } \right) \\
 \quad \quad = (144p - 384)\times \frac{\textstyle M^2}{\textstyle p^2}\times t_{sendrec} +
(\frac{\textstyle 3}{\textstyle 4}p + 6)\times t_{delay} +
\frac{\textstyle 24}{\textstyle p}\times \left( {M^2\log _2
\left( {M^2} \right)\;t_c } \right). \\
 \end{array}
\end{equation}

\end{itemize}

For all the PTF discussed above (Eq. (26)-(29)), it is clear that
$T_{delay} $ will take larger portion of $T_{sum} $ when $p$
becomes larger in those schemes. For fixed $p$, $T_{delay} $ will
take smaller portion of $T_{sum} $ when more groups are adopted
in the PTF schemes (e.g. 4-N or 8-N scheme). Hence, the 4-N and
8-N schemes have some advantage over the 1-N and 2-N schemes when
$p$ is large. Some further discussions for Eqs. (26) - (29) will
be continued in the next subsection together with the speedup
plots (Figs.7).

Like the 2D NS equations (Eqs. (4) and (5)), Eq. (24) and (25) can
also be written in the conservative form:
\begin{eqnarray}
&& \rho _t + \nabla \cdot \left( {\rho {\rm {\bf u}}}
\right) = 0,
\label{eq30}
\\
&& \left( {\rho \omega } \right)_t + \nabla \cdot
\left( {\rho \omega {\rm {\bf u}}} \right) = - \frac{1}{2}\left(
{\rho ^2} \right)_x.
\label{eq31}
\end{eqnarray}
When the Fourier-Galerkin spectral methods are used to solve the
equations above, Eq. (31) presents some problem because the time
evolution step is carried out in spectral space: to get the value
of $\hat {\omega }$ in the next time step, $\widehat{\left( {\rho
\omega } \right)}$ has to be transferred back to physical space so
the new value can be obtained by using the relation $\omega =
{\left( {\rho \omega } \right)} \mathord{\left/ {\vphantom
{{\left( {\rho \omega } \right)} \rho }} \right.
\kern-\nulldelimiterspace} \rho $. Furthermore, there is an extra
nonlinear term $ - \frac{1}{2}\left( {\rho ^2} \right)_x $ in Eq.
(33) which requires one extra FFT. The total number of FFTs in
conservative form is the same as non-conservative form. Hence,
unlike what we did for 2D NS equations in section 2, the
non-conservative equations are solved here.

\subsection{Benchmarks and comparisons}

In this subsection, we will show the speedup plots of the
Boussinesq equations on SGI3800. Unlike what we did for the NS
equations, we added a new resolution ($64^2$) to show the
effectiveness of the new parallel scheme in the low resolution.
The maximum processors used here is 32, which makes the resulting
speedup plots (Figs. 7) more concise. It is also easier to use
less than 32 processors in our 64-nodes machine; because we have
to shut all other computing jobs down if we want to do the 64
CPUs run.

\begin{figure}[t!]
\centering
\begin{minipage}[c]{.4 \linewidth}
\scalebox{1}[1]{\includegraphics[width=\linewidth]{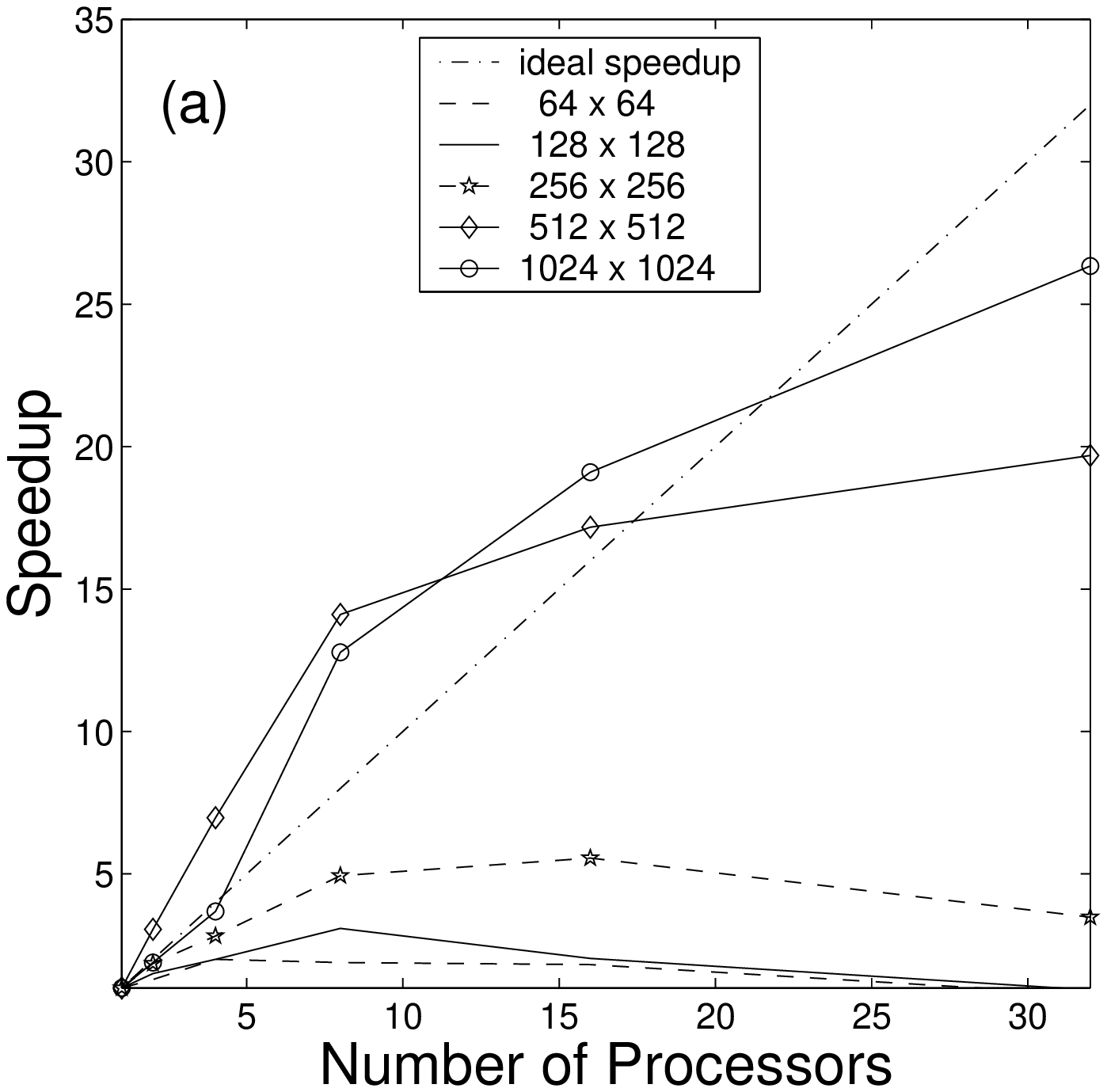}}
\end{minipage}
\begin{minipage}[c]{.4 \linewidth}
\scalebox{1}[1]{\includegraphics[width=\linewidth]{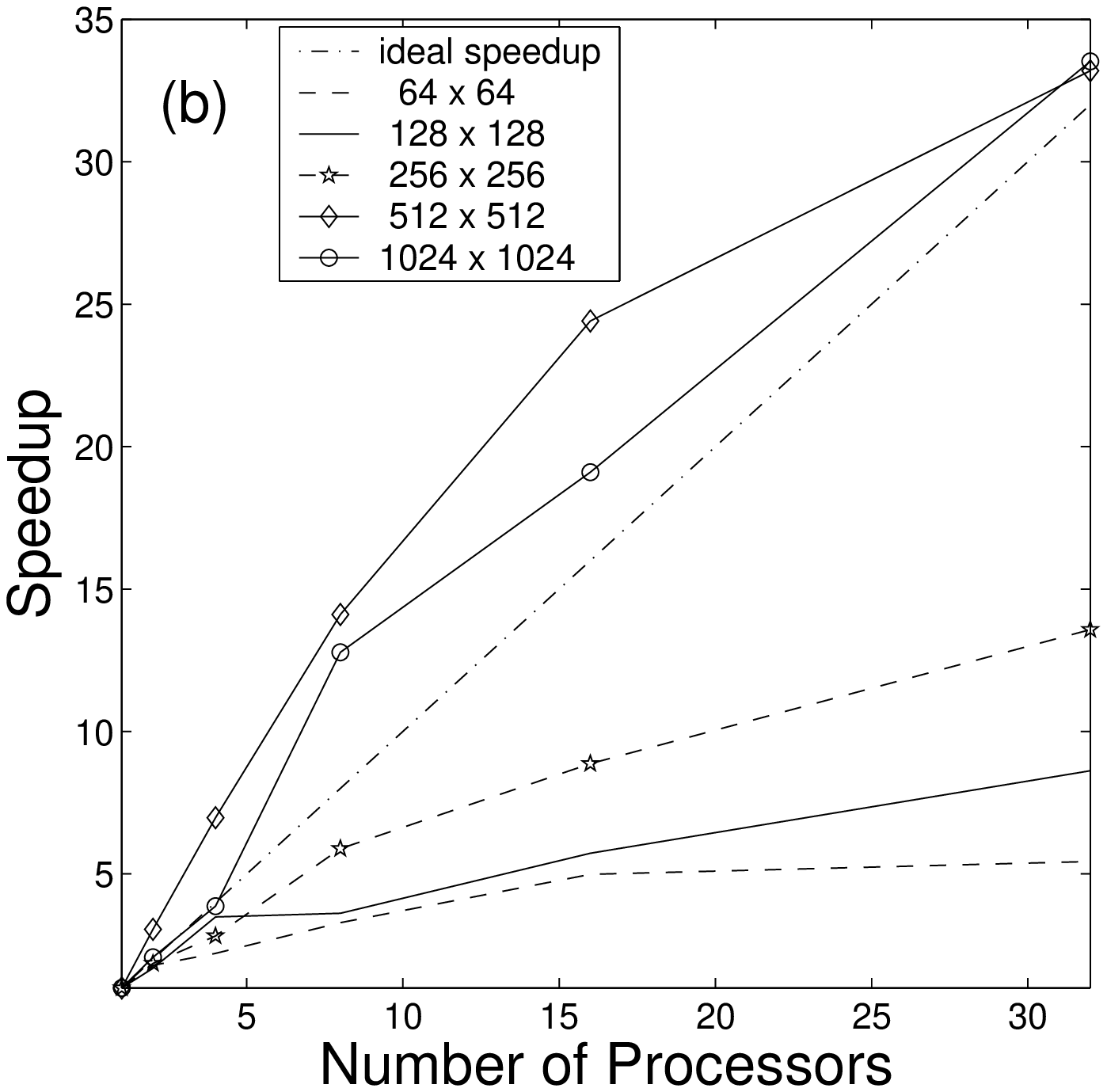}}
\end{minipage}
\caption{(a) The speedup of 2D Boussinesq code with PFFT scheme
(or ``1-N'' scheme). (b) The speedup of 2D Boussinesq code with
``ideal'' scheme. The values used on (b) are the top speedup
among 1-N, 2-N, 4-N, and 8-N scheme for same resolutions and same
numbers of processors. }
\end{figure}

Fig. 7 (a) shows the speedup plot for PFFT (or. 1-N) scheme. The
top speedup is obtained on 4 processors for the $64^2$ grid,
while 8 and 16 processors reach the top speed for the resolutions
of $128^2$ and $256^2$, respectively. The speedup curves of
$512^2$ and $1024^2$ show reasonably good parallel efficiency of
PFFT scheme. The $512^2$ runs show super linear speedup for 2, 4,
8, and 16 nodes, which is due to the cache effect. The $1024^2$
runs only have super linear speedup in the cases of 8 and 16
nodes. The speedups of 32-nodes run for these two high
resolutions are lower than 32 because $T_{delay} $ begins to
dominate the total computation time.

We did not show the speedup plots for resolutions higher than
$1024^2$ because of the limit of the maximum local memory.
However, since solving the equations with the highest possible
resolution is the main task of our research, we will discuss the
parallel efficiency on those high resolutions ($2048^2$ and
$4096^2$) in other ways later in the next subsection.

Fig. 7(b) shows the ``best speedup'' plot of PTF schemes. All the
resolutions have their top speed in the 32 CPUs case. Moreover,
almost all the points on the curves show super linear speedup for
higher resolutions ($512^2$ and $1024^2$).

The peak speeds shown in Fig. 7(b) are increased by a factor of
27{\%} ($1024^2$ resolution) to 171{\%} ($64^2$ resolution)
compared to the PFFT scheme. The lower resolutions get a higher
factor because the limitation of the total CPUs available. We can
predict that the factor of $1024^2$ resolution will be larger
than 27{\%} if the maximum number of CPUs used is larger than 32.

\begin{table}[t!]
\caption{The corresponding schemes for the values in Fig. 7(b).}
\centering
\begin{tabular}
{|p{53pt}|p{33pt}|p{33pt}|p{33pt}|p{33pt}|p{33pt}|} \hline\hline
\textbf{}& $64^2$\textbf{}& $128^2$\textbf{}& $256^2$\textbf{}&
$512^2$\textbf{}&
$1024^2$\textbf{} \\
\hline 1 CPU& 1-N& 1-N& 1-N& 1-N&
1-N \\
\hline 2 CPUs& 2-N& 2-N& 1-N& 1-N&
2-N \\
\hline 4 CPUs& 2-N& 2-N& 1-N& 1-N&
2-N \\
\hline 8 CPUs& 8-N& 2-N& 2-N& 1-N&
1-N \\
\hline 16 CPUs& 8-N& 8-N& 2-N& 2-N&
1-N \\
\hline 32 CPUs& 8-N& 8-N& 8-N& 4-N&
2-N \\
\hline \hline
\end{tabular}
\end{table}

Table 5 shows the corresponding schemes to the points on the Fig.
7(b). PFFT (or, 1-N) scheme never show the best performance for
resolutions of $64^2$ and $128^2$ for $p \ge 2$ (For $p = 1$, 1-N
is the only choice, which is not necessary for comparison). For
resolutions higher than $256^2$, the 1-N scheme reaches the peak
speed more frequently, especially for runs with fewer processors.

As indicated in Table 5, the PTF schemes with more groups (e.g.
4-N or 8-N scheme) have better speedup for lower resolutions and
larger number of CPUs, while the schemes divided into fewer
groups (e.g. 1-N or 2-N scheme) work best for higher resolutions
and relatively smaller number of CPUs.

In the real programming efforts, we found that it is more
convenient to fix the size of each group (instead of fixing $p$)
because the size of main calculating arrays can be determined
beforehand. The code will get the number of groups (1, 2, 4, or
8) at the initial stage of real runs, and distribute the 20 FFTs
into different groups. Thus, the programming effort concerning
the PFT scheme is trivial if the PFFT code is already available.

Attention should be drawn to the performance of the 2-N scheme,
which occupy eight of the total thirty places in Table 5. In the
$1024^2$ runs, the 2-N scheme is faster than the 1-N scheme on 2
and 4 nodes' simulations, which is counter to the conclusion we
made in the above paragraph. Although most of the weird speedup
behaviors in parallel computing can be attributed to cache
effect, this one presents some difficulties because normally the
more CPUs are used to calculate one FFT, the smaller $t_c $ will
be (see the discussion in Section 2.2). To explain this, we need
to calculate the exact values of $T_{comm} $ for the 1-N and 2-N
schemes in our analytical models (Eqs. (26) and (27)).

For the 1-N scheme,

\[
T_{comm} = \left\{ {{\begin{array}{*{20}c}
 {10M^2\times t_{sendrec} + 40\times t_{delay} \quad \quad \mbox{if}\;p = 2}
\hfill \\
 {7.5M^2\times t_{sendrec} + 120\times t_{delay} \quad \,\mbox{if}\;p = 4.}
\hfill \\
\end{array} }} \right.
\]

For the 2-N scheme,

\[
T_{comm} = \left\{ {{\begin{array}{*{20}c}
 {4M^2\times t_{sendrev} + 4\times t_{delay} \quad \quad \mbox{if}\;p = 2}
\hfill \\
 {7M^2\times t_{sendrev} + 24\times t_{delay} \quad \;\,\mbox{if}\;p = 4.}
\hfill \\
\end{array} }} \right.
\]

It is clear that all parts of $T_{comm} $ in the 2-N scheme are
smaller than the 1-N scheme in the case of $p \le 4$, although
$T_{sendrec} $ of the 2-N scheme is larger than that of the 1-N
scheme for $p \ge 8$. Again, we assume that $t_c $ remains
roughly the same for different schemes. The complete explanation
has to take ``cache effect'' into consideration.

\subsection{ Numerical results for the Boussinesq convection}

\begin{figure}[t!]
\centering
\begin{minipage}[c]{.5 \linewidth}
\scalebox{1}[1]{\includegraphics[width=\linewidth]{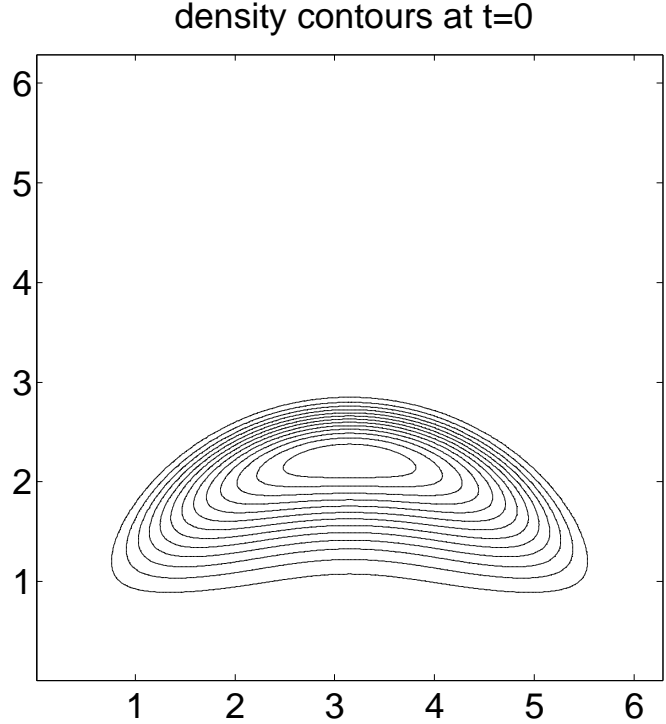}}
\end{minipage}
\caption{The contour plot of initial density, with the resolution
of $1024^2$.}
\end{figure}

In this subsection, we will show some numerical simulations
calculated by our parallel codes. For ease of comparison with
former results, we adopted the same initial condition in Ref
~\cite{we94}:

\begin{equation}
\label{eq32} \omega (x,y,0) = 0,
\end{equation}

\begin{equation}
\label{eq33} \rho (x,y,0) = 50\rho _1 (x,y)\rho _2 (x,y)\left[ {1
- \rho _1 (x,y)} \right],
\end{equation}
where
\[
\rho _1 (x,y) = \left\{ {{\begin{array}{*{20}c}
 {\exp \left( {1 - \frac{\textstyle \pi ^2}{\textstyle \pi ^2 - x^2 - (y - \pi )^2}} \right),\quad
\mbox{if}\;x^2 - (y - \pi )^2 \le \pi ^2,} \hfill \\
 {0,\quad \quad \quad \mbox{otherwise,}} \hfill \\
\end{array} }} \right.
\]

\[
\rho _2 (x,y) = \left\{ {{\begin{array}{*{20}c}
 {\exp \left( {1 - \frac{\textstyle \left( {1.95\pi } \right)^2}{\textstyle \left( {1.95\pi }
\right)^2 - (x - 2\pi )^2}} \right),\quad \mbox{if}\;\left| {x -
2\pi }
\right| \le 1.95\pi, } \hfill \\
 {0,\quad \quad \quad \mbox{otherwise.}} \hfill \\
\end{array} }} \right.
\]

Fig. 8 shows the initial contour of $\rho $ on a grid of
$1024^2$. The cap-like contour will develop into a rising bubble
during the evolution, with the edge of the cap rolling up.

\begin{figure}[t!]
\centering
\begin{minipage}[c]{.4 \linewidth}
\scalebox{1}[1.0]{\includegraphics[width=\linewidth]{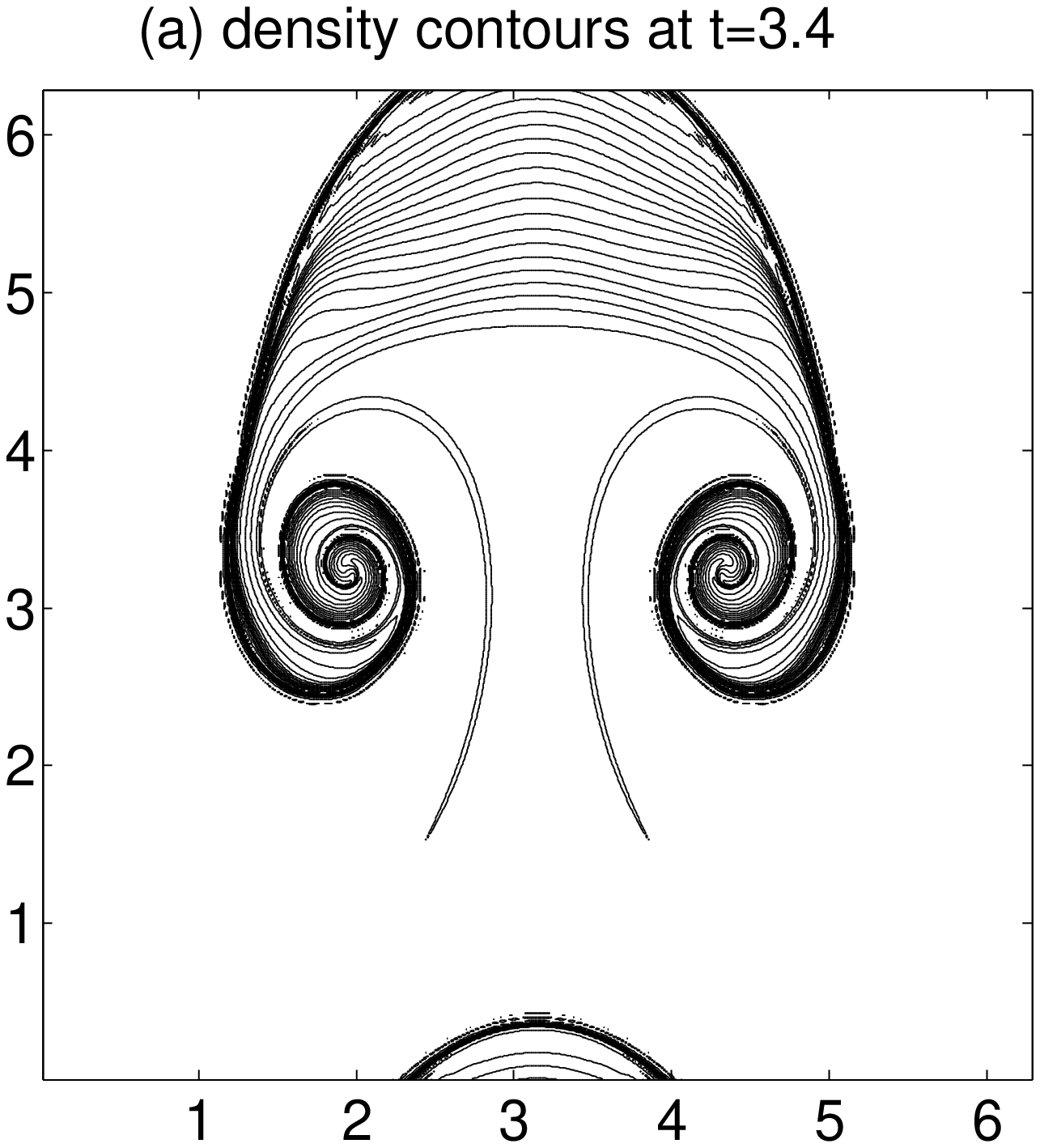}}
\end{minipage}
\begin{minipage}[c]{.4 \linewidth}
\scalebox{1}[1.0]{\includegraphics[width=\linewidth]{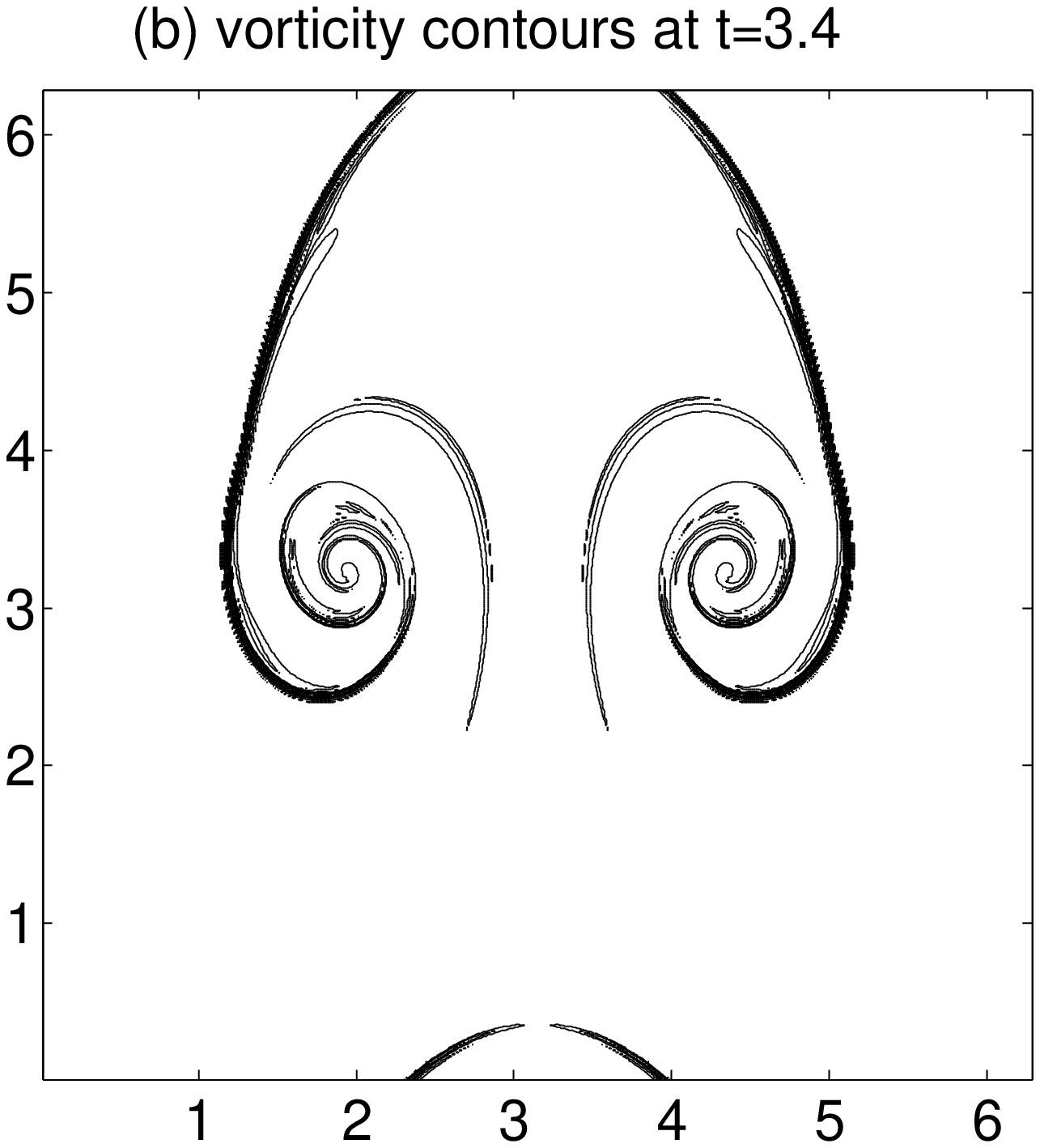}}
\end{minipage}
\begin{minipage}[c]{.4 \linewidth}
\scalebox{1}[1.0]{\includegraphics[width=\linewidth]{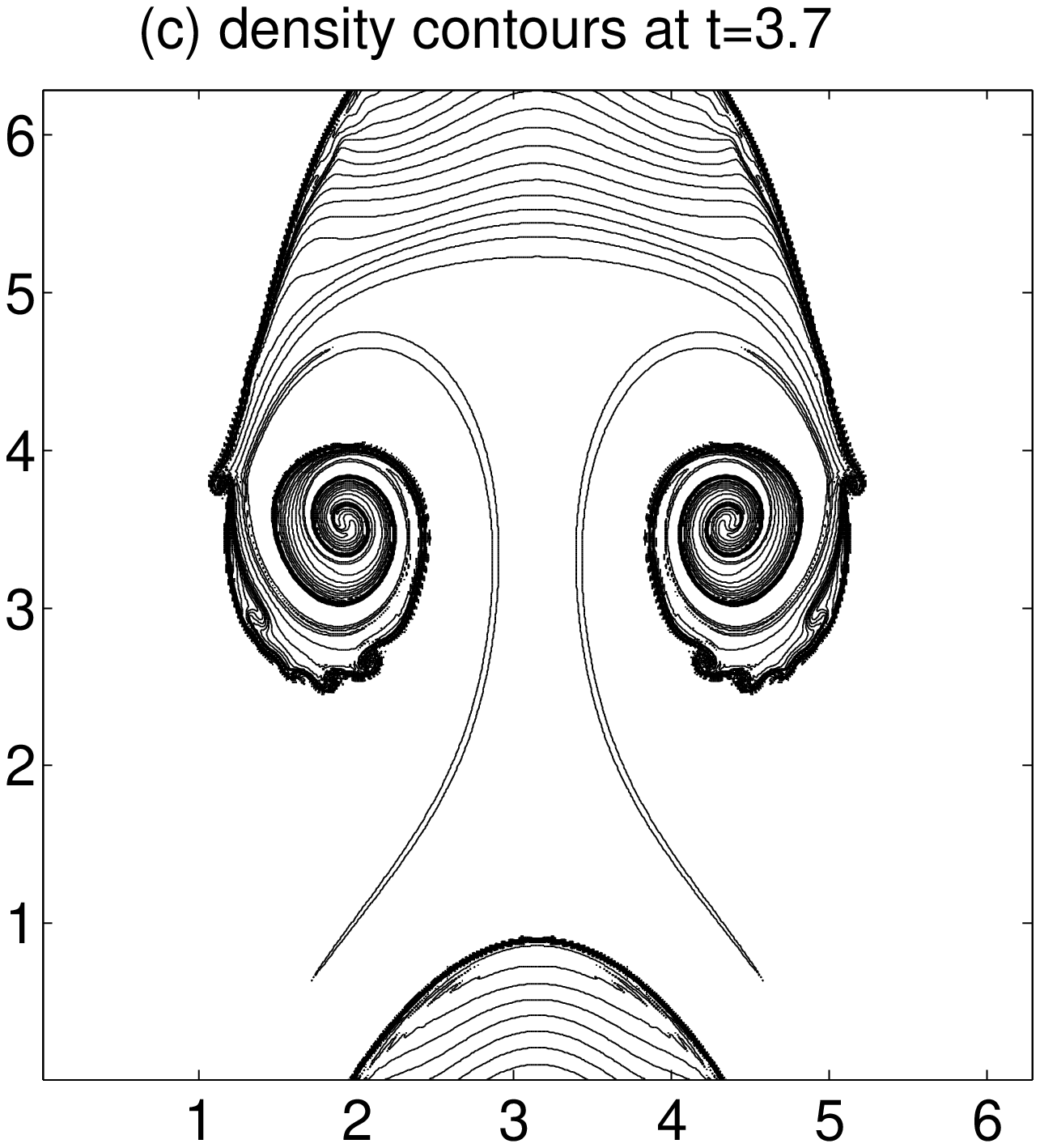}}
\end{minipage}
\begin{minipage}[c]{.4 \linewidth}
\scalebox{1}[1.0]{\includegraphics[width=\linewidth]{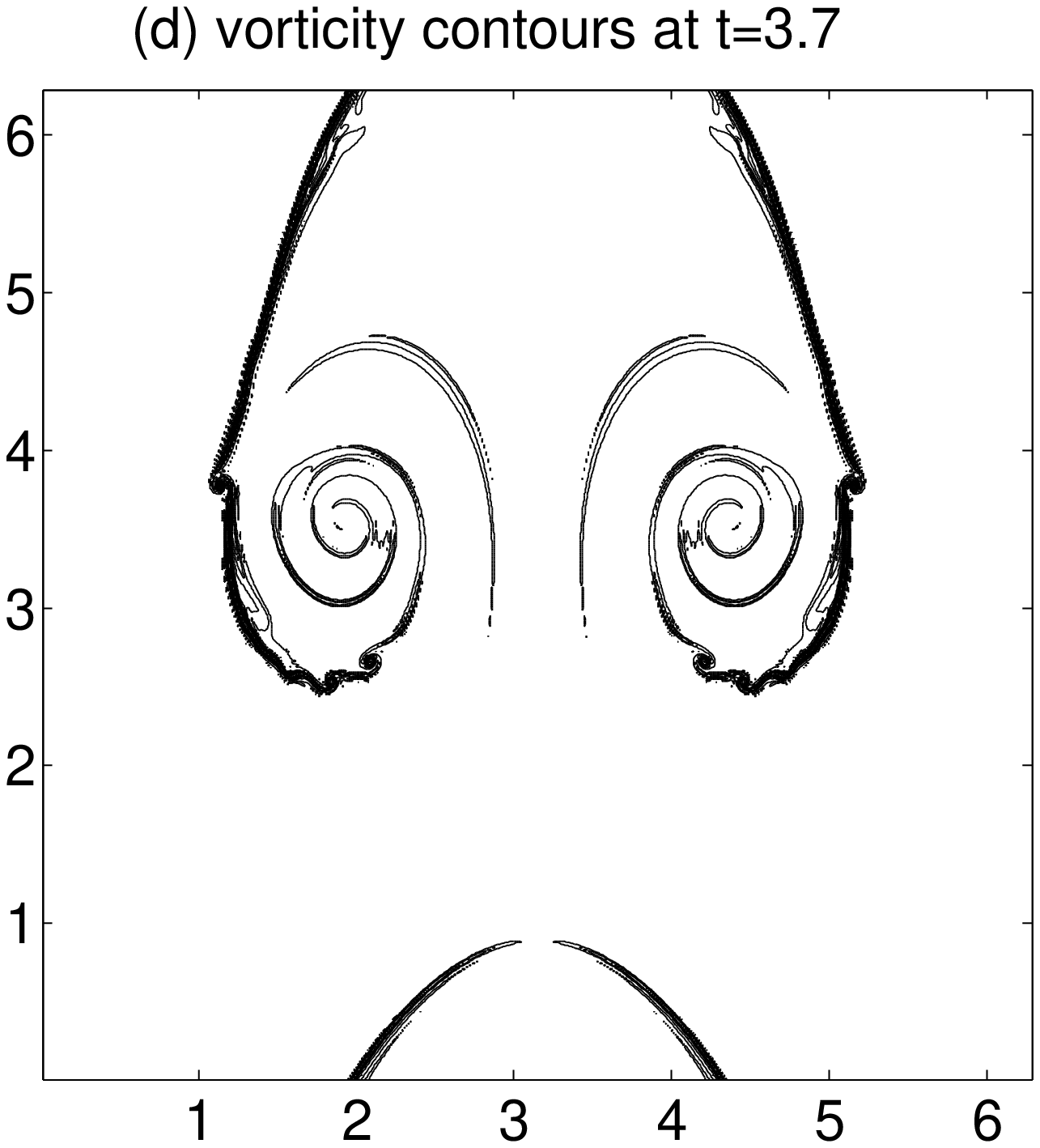}}
\end{minipage}
\caption{(a) and (b) show the contour plots of density and
vorticity at t=3.4 with the resolution of $1024^2$, (c) and (d)
show the contour plots at t=3.7.}
\end{figure}

\begin{figure}[t!]
\centering
\begin{minipage}[c]{.4 \linewidth}
\scalebox{1}[1.0]{\includegraphics[width=\linewidth]{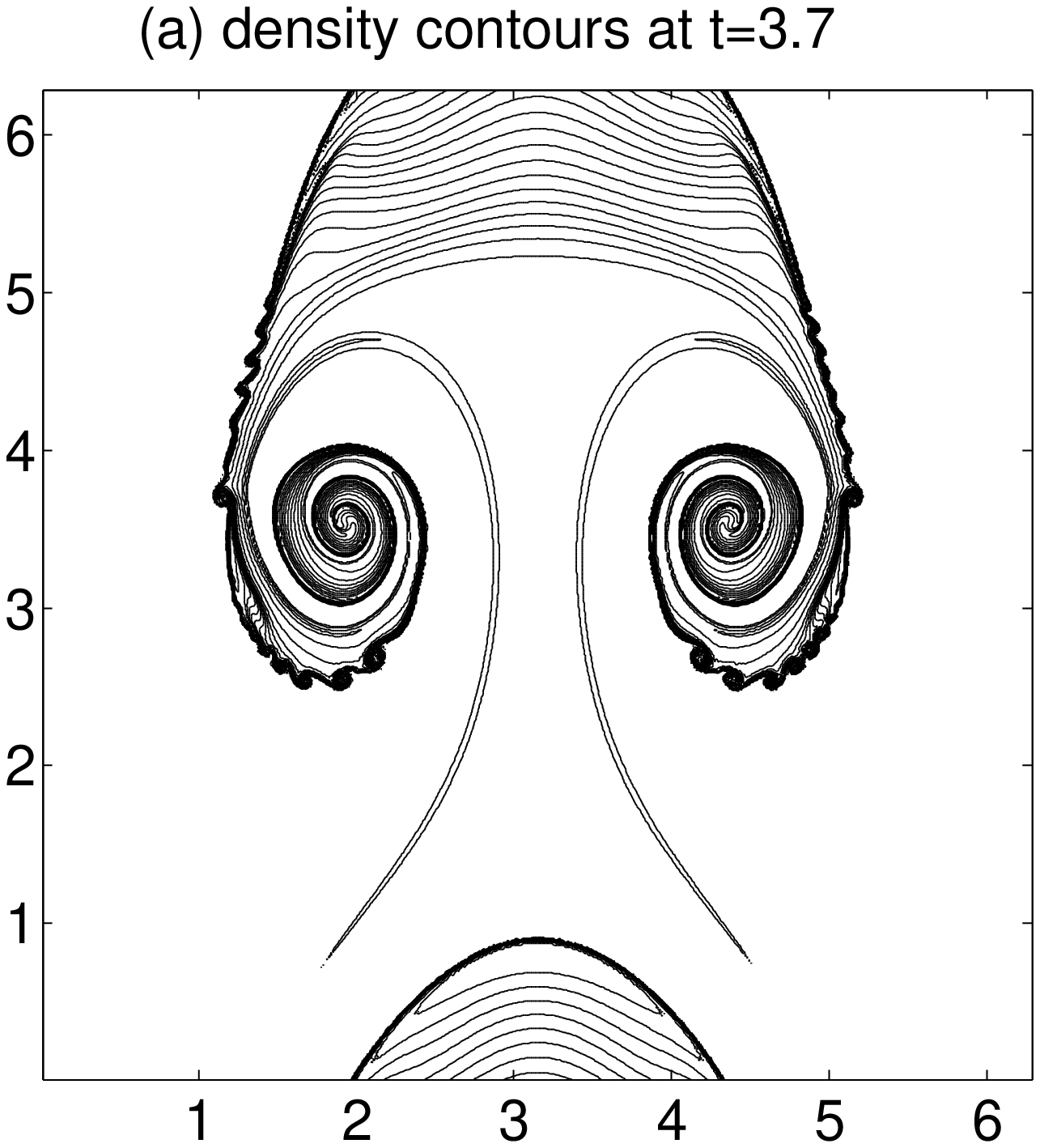}}
\end{minipage}
\begin{minipage}[c]{.4 \linewidth}
\scalebox{1}[1.0]{\includegraphics[width=\linewidth]{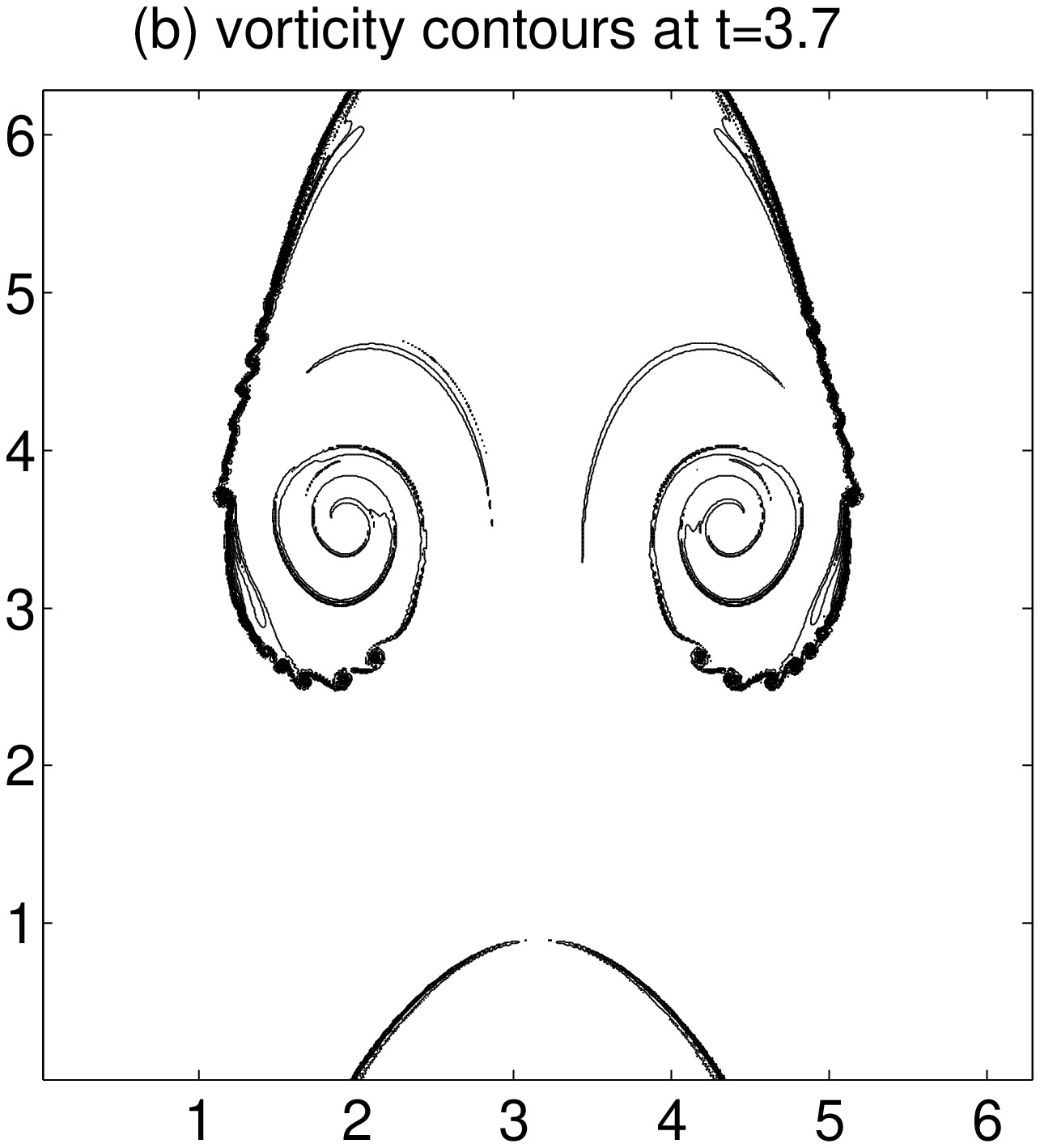}}
\end{minipage}
\begin{minipage}[c]{.4 \linewidth}
\scalebox{1}[1.0]{\includegraphics[width=\linewidth]{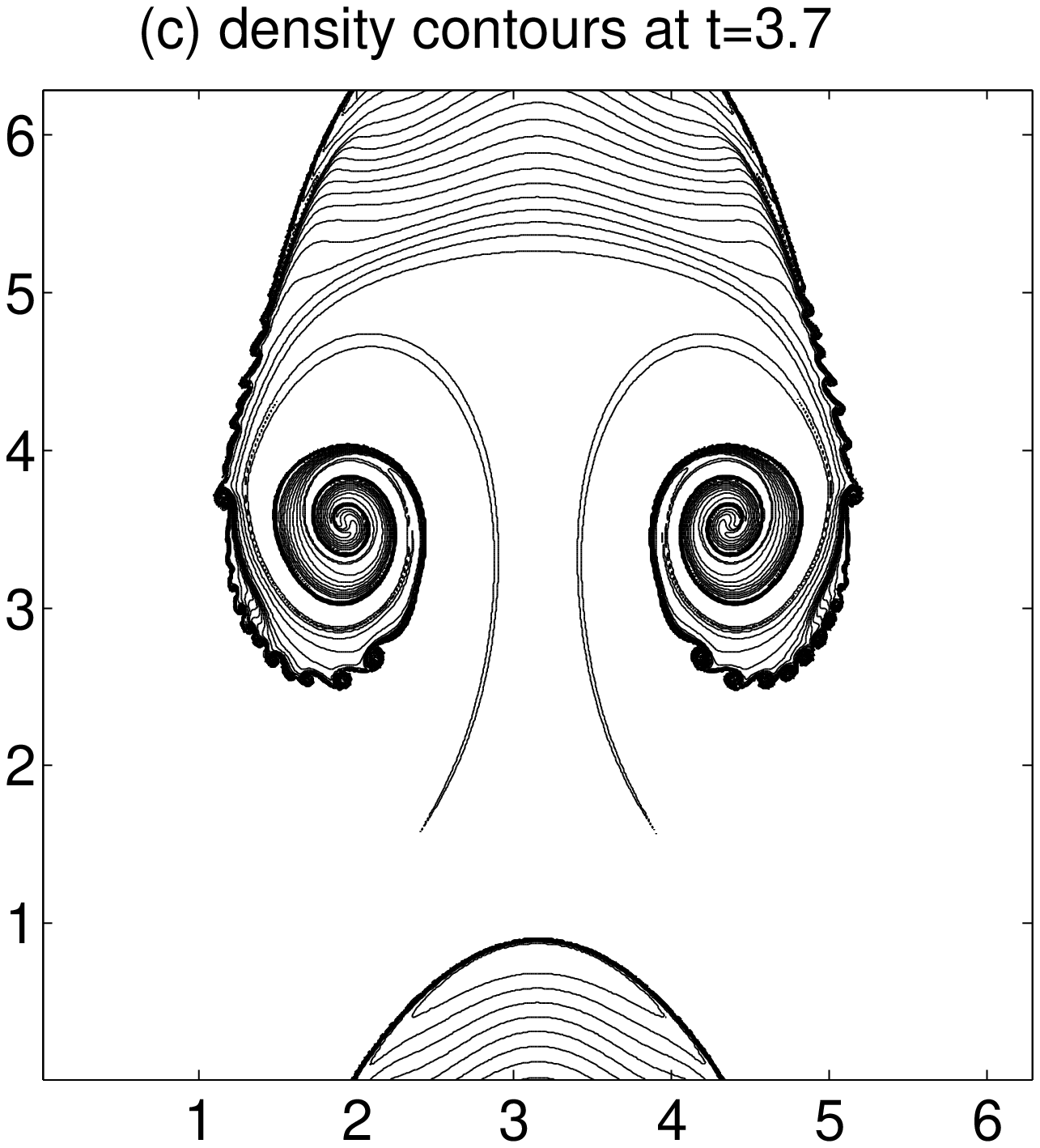}}
\end{minipage}
\begin{minipage}[c]{.4 \linewidth}
\scalebox{1}[1.0]{\includegraphics[width=\linewidth]{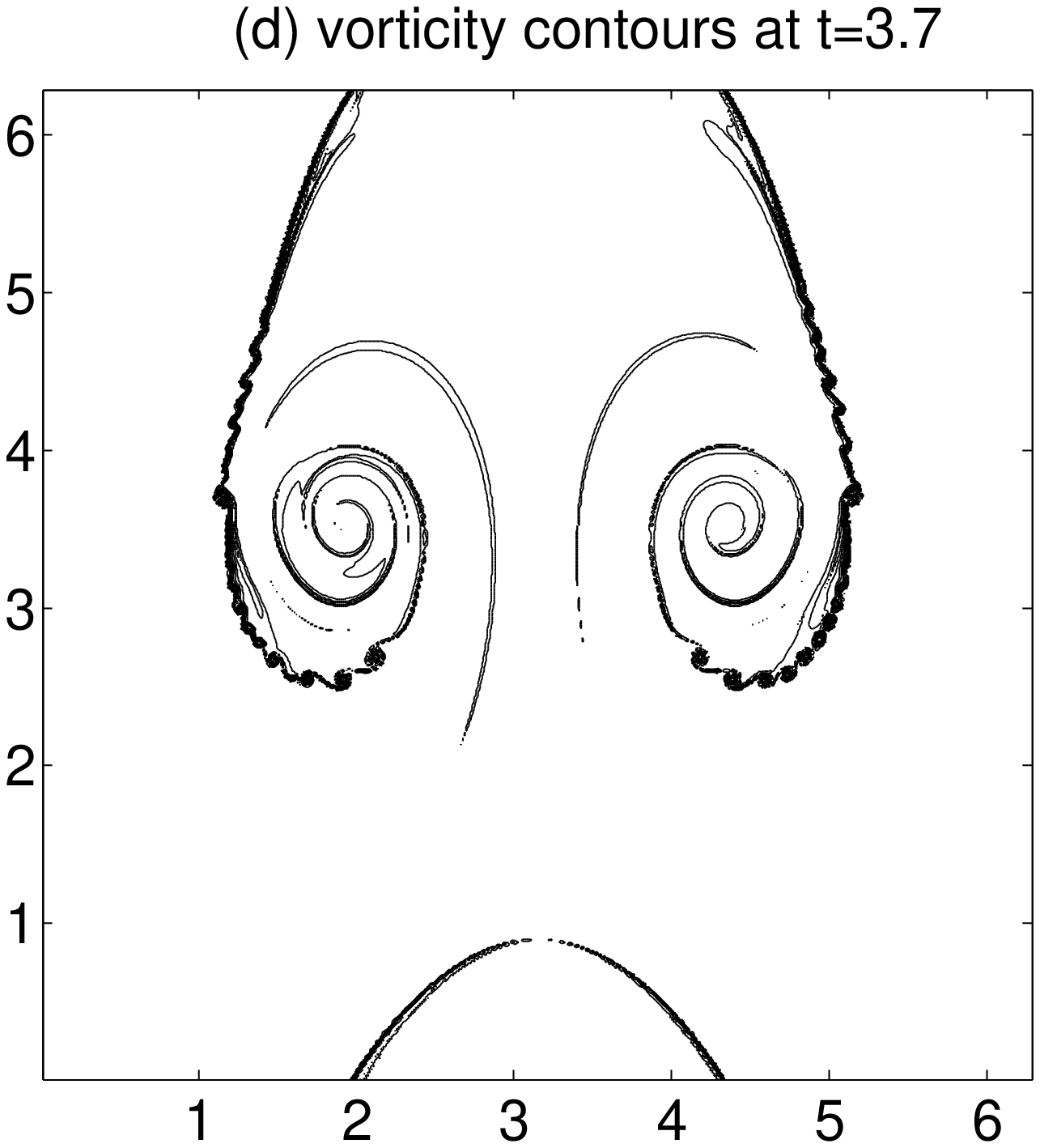}}
\end{minipage}
\caption{(a) and (b) are the contour plots of density and
vorticity at t=3.7, with the resolution of $2048^2$. (c) and (d)
are the contour plots with the resolution of $4096^2$ at t=3.7.}
\end{figure}

At $t \simeq 3.4$, the density and vorticity contour develop into
the shape of ``two eyes.'' Figs. 9 (a) and (b) show the results
with the resolution of $1024^2$. For spectral methods with
different resolutions ($2048^2$ run of our simulation, and
$1500^2$ run of ~\cite{we94}) and ENO finite difference methods
in ~\cite{we94}, some good agreements have been observed.

Arguments arise when the computations are carried on. At $t \simeq
3.7$, the smooth edge of the rolling eyes becomes unstable, see
Figs. 9(c) and (d). Whether this phenomenon is physically real or
not is still an open question. It was argued that this collapse is
due to the numerical effect ~\cite{we94,ceni01}.

As discussed above, the most straightforward way to investigate
this problem is to increase the resolution of simulations.
Therefore, we also adopted grids of $2048^2$ and $4096^2$ in the
simulations.

For the $1024^2$ run, the time step used is 0.0008 which is
determined by the CFL condition. It took the parallel 2-N scheme
twenty hours to reach $t = 3.7$ with 32 processors. It takes 25
hours for PFFT schemes, and 28 days for the serial code.

For the $2048^2$ run, the time step used is 0.0004. The speed of
the 1-N scheme and the 2-N scheme with 32 nodes are very close (a
difference of 2{\%}). We use the 1-N scheme because it is
slightly faster. The code lasts for 13 days before it reaches $t =
3.7$. For serial code, it may take one year provided the memory of
the computer is large enough.

For the $4096^2$ run, the time step used is 0.0002. The 1-N scheme
is the only choice because other schemes require larger computer
memory: in principle, the memory size of the 2-N, 4-N, and 8-N
schemes are 2, 4, and 8 times as large as that of the 1-N scheme.
For this run, it is even a burdensome job for the parallel
computation with 32 processors, which may last for four months in
total. We actually use the interpolated data of the $2048^2$ run
at $t = 3.4$ as the initial condition. It is believed that the
solution at $t = 3.4$ is smooth. The code lasts for three weeks
before it reaches $t = 3.7$.

Figs. 10(a, b) show contour plots of the $2048^2$ run at $t =
3.7$, on which the collapse is observed in the smooth edge; the
same phenomenon is observed in the $4096^2$ run (Figs. 10(c, d)).
It is noticed that more rolling vortices are observed on the edge
of the ``eyes'' in Fig. 10(a) than that in Fig. 9(c), and even
more vortices appear in Fig. 10(c). There are some trivial
differences if we make a detailed comparison within these three
runs, but Figs. 9(c, d) and Figs. 10 clearly show that the
collapsing process is not a non-physical phenomenon from the
numerical artifacts. If we started the $4096^2$ run from $t = 0$
using Eq. (32) and (33) to get the initial data, the contour plots
might be slightly different from what we obtained in Figs. 10(c,
d). The main task here is to determine the possibility of the
collapse phenomenon, the potential small difference in the flow
pattern will not hurt the general conclusion. Some further
details of the physical phenomenon will be discussed in another
work. Here, we will try to focus our topic on the parallel solver.

Unlike the simulations in Section 2, most of computations (two of
the three) in this section use the traditional PFFT scheme mainly
due to the limitation of the maximum CPU number. If more
processors are available so that the computer memory presents no
problem to the PTF scheme, then the speed of PTF will also be
faster than PFFT.

\section{Conclusions and discussions}

To sum up, the PTF scheme takes the advantages of the PTD and PFFT
schemes, and can increase peak speed of codes significantly. The
PTF scheme works best for low resolution run, or relatively large
resolution with large number of CPUs. For very high resolutions,
the PTF scheme is most likely slower than PFFT if not many
processors are involved; PTF also needs more memory than that for
PFFT, which presents some limitations on the scale of the highest
resolution that one particular parallel computer can handle.
These two disadvantages of PTF explain why similar schemes are not
very popular in the 3D simulations where the array sizes may be
even larger than that for the 2D simulations with high
resolutions. It is advisable to treat PFFT as a special case of
PTF, and the combined version of the parallel codes will always
yield the best performance after a small amount of preparing work
(e.g. making a table similar to Table 2 and Table 5 on the
particular parallel machine to be used). In the meanwhile, the
simple analytical mode of the parallel scheme, which divide
$T_{sum} $ into three parts: $T_{sendrec} $, $T_{delay} $, and
$T_{comp} $, is useful in explaining the performance of the
parallel codes.

The parallel code of the NS equations helps us to find another
example of double-valued $\omega$-$\psi $ structure; and the
parallel Boussinesq code reveals the insight of an open question,
which suggests that the collapse of the bubble cap may be a
candidate for the finite time singularity formation. Both sets of
investigation show that the parallel computing is useful in large
scale numerical simulations.

We conclude our paper by some discussions on the future parallel
computing. It is well-known that the speed of single processor
increases much faster than the speed of memory access (five times
faster in the last decade). Hence the latency of memory access in
terms of processor clock cycle grows by a factor of five in the
last ten years ~\cite{cull99}. For a fixed type of parallel
scheme, the speedup curves (see Figs. 1-3 and 7) for high
resolution (e.g. $1024^2$) in future machines may be similar to
that of low resolutions in current computers ($64^2$ or $128^2$).
In the case of PFFT scheme applied on a $1024^2$ resolution, it is
possible that the speed of 64-nodes run is slower than that of
the serial code in the future. On the other hand, the PTF scheme
which uses much less $T_{delay} $ than that of the PFFT scheme
has greater advantage over the PFFT scheme in parallel computing.
In other words, the PTF strategy is also a scheme for the future.

\vskip .25cm
\noindent
{\bf Acknowledgments.}
We would like to thank Linbo Zhang, Zhongze Li, and Ying Bai for
the support of using local parallel computers. ZY also thanks
Prof. W.H. Matthaeus who supplied the original serial FORTRAN 77
Navier-Stokes pseudospectral codes. TT thanks the supports from
International Research Team on Complex System, Chinese Academy of
Sciences, and from Hong Kong Research Grant Council.

\end{document}